\documentclass{aa}

\usepackage{graphicx}
 
   \def\tick{
    \begin{picture}(18,26)
    \thicklines
    \put(0,10){\line(1,-2){5}}
    \put(5,0){\line(1,2){13}}
    \end{picture}
    }
    
\input psfig.tex  
  
\begin{document} 

\title{The Master Catalogue of stars towards
        the Magellanic Clouds\thanks{This work has been supported by {\sc astrovirtel}, a project
	funded by the European Commission under FP5 Contract No. HPRI-CT-1999-00081.}} 
   \titlerunning{Master Catalogue of stars towards the Magellanic Clouds.}

\subtitle{I. Multispectral surveys of the Large Magellanic Cloud}

\author{N. Delmotte
  \inst{1,4}
  \and
   C. Loup
   \inst{2}
   \and 
   D. Egret
  \inst{1}
  \and
   M.-R. Cioni
   \inst{3,4} 
   \and 
   F. Pierfederici
  \inst{4}
          }
          
   \offprints{N. Delmotte}
          
\institute{CDS, Observatoire Astronomique de Strasbourg, UMR 7550, 
    Universit\'e Louis Pasteur, 67000 Strasbourg, France \\
    \email{delmotte@astro.u-strasbg.fr, egret@astro.u-strasbg.fr}
\and
    Institut d'Astrophysique de Paris, CNRS UPR 341, 98 bis Bd Arago, 75014 Paris, France \\
    \email{loup@iap.fr}
\and
    Leiden Observatory, University of Leiden, P.O. Box 9513, 2300 RA Leiden, The Netherlands 
\and
    European Southern Observatory, ESO, K.-Schwarzschild-Str.-2, D-85748 Garching, Germany \\
    \email{mcioni@eso.org}
   }

\date{Received 28 March 2002 ; Accepted 05 September 2002}

\abstract{
The Master Catalogue of stars towards the Magellanic Clouds (MC2)\thanks{\tt http://vizier.u-strasbg.fr/MC2/}
is a multi--wavelength reference catalogue. The current paper presents the first results of the MC2 project. 
We started with a massive cross--identification
of the two recently released near--infrared surveys:
the DENIS Catalogue towards the
Magellanic Clouds (DCMC) with more than 1.3 million sources identified in at least two
of the three DENIS filters ($I$ $J$ $K_\textrm{s}$) and  the
2nd Incremental Release of the 2MASS
point source catalogue ($J$ $H$ $K_\textrm{s}$) covering the same region of the
sky. Both point source catalogues provide an unprecedented wealth of data on the stellar
populations of the Magellanic Clouds (MCs). The cross--matching procedure has been extended to optical wavelength ranges, including 
the UCAC1 (USNO) and GSC2.2 catalogues.
New cross--matching 
procedures for very large catalogues have been developed and important results on the astrometric and photometric
accuracy of the cross--identified catalogues were derived. The cross--matching of 
large surveys is an essential tool to improve our understanding of their 
specific contents. 
 This study has been partly supported by the 
 {\sc astrovirtel}\thanks{\tt http://www.stecf.org/astrovirtel/} project that
aims at improving access to astronomical archives as virtual telescopes.
}

\maketitle


\section{Introduction}

\begin{figure*}
\begin{center}
   \begin{tabular}{cc}
 	  \psfig{figure=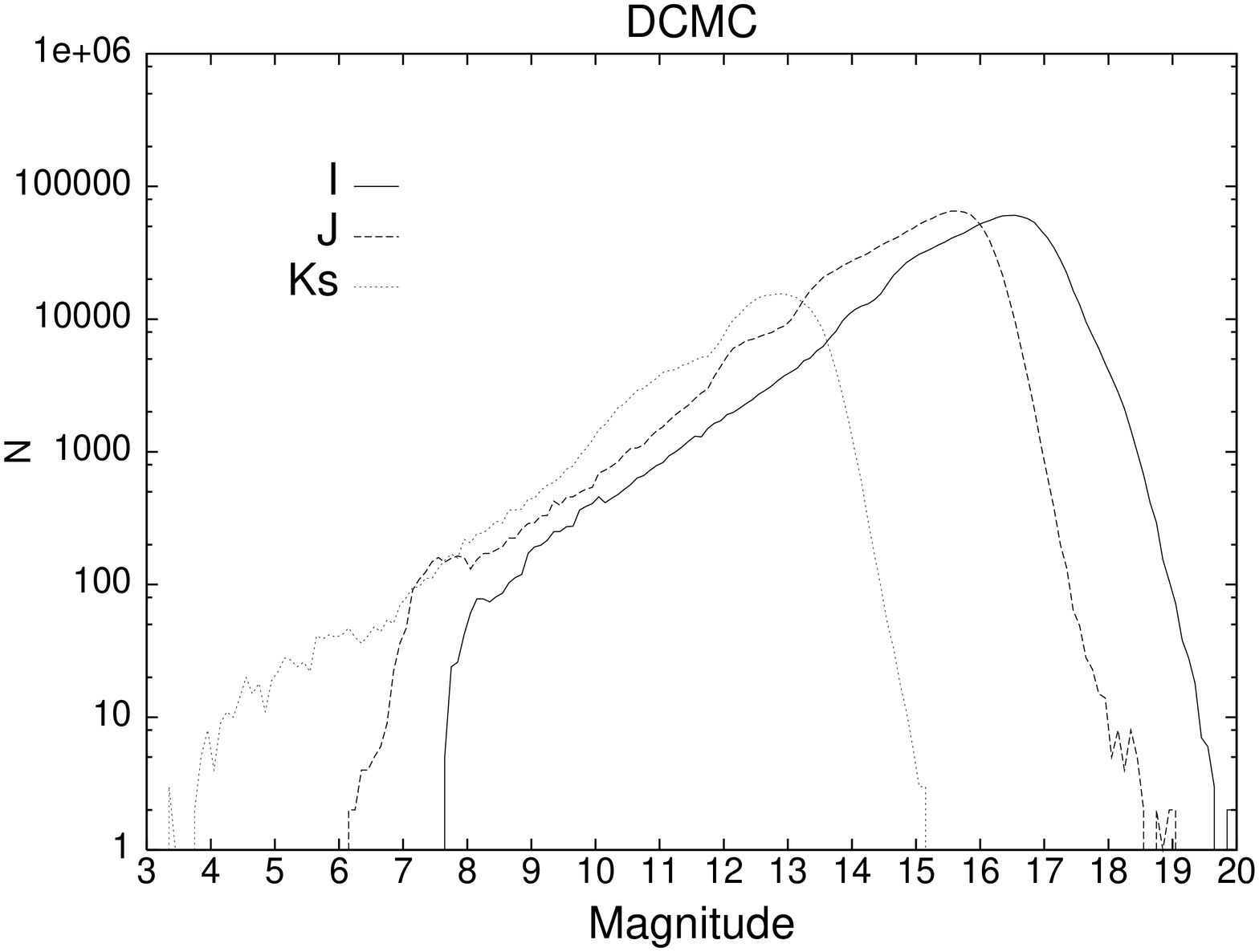,clip=,height=5cm,angle=-90} & \psfig{figure=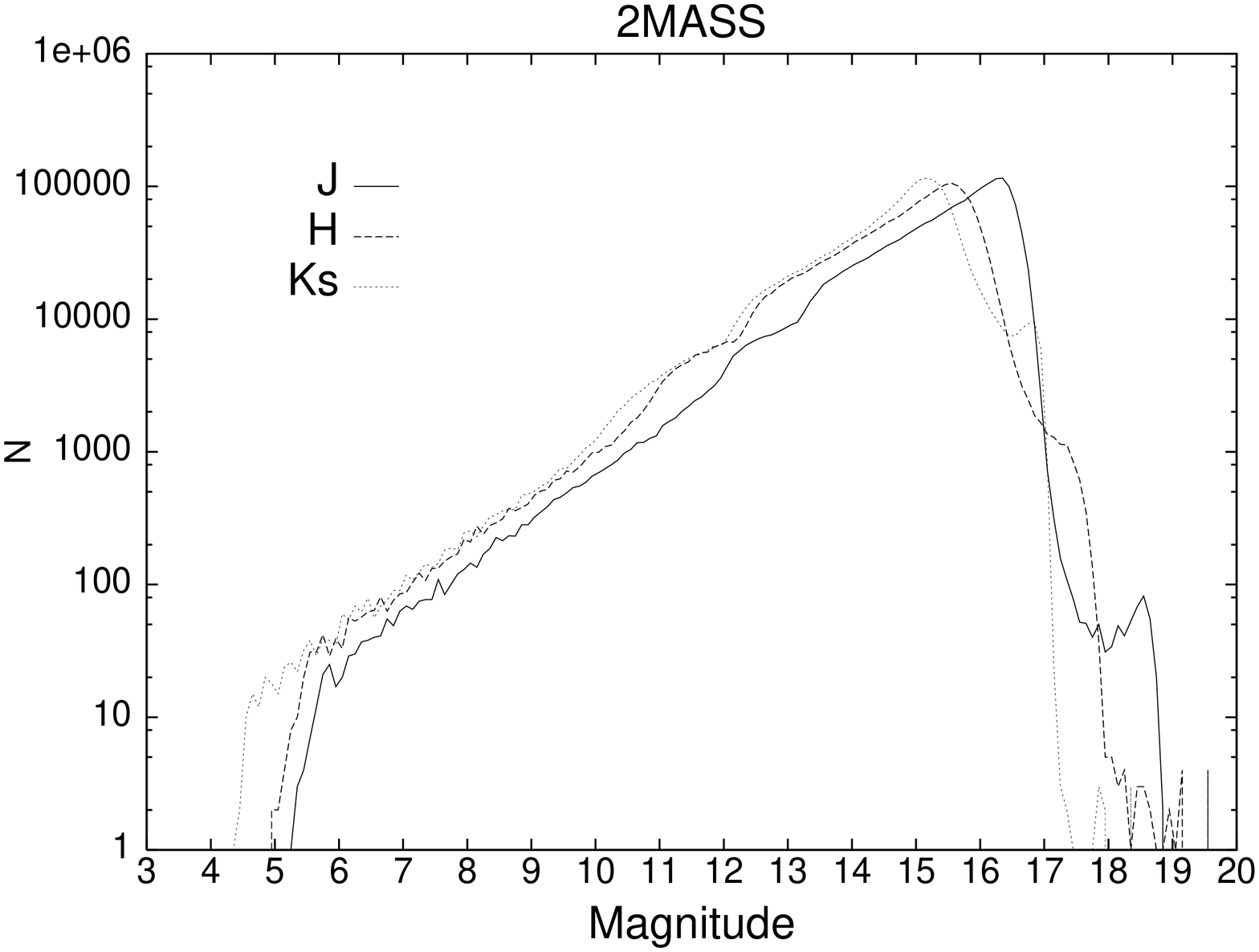,clip=,height=5cm,angle=-90} \\
 	  \psfig{figure=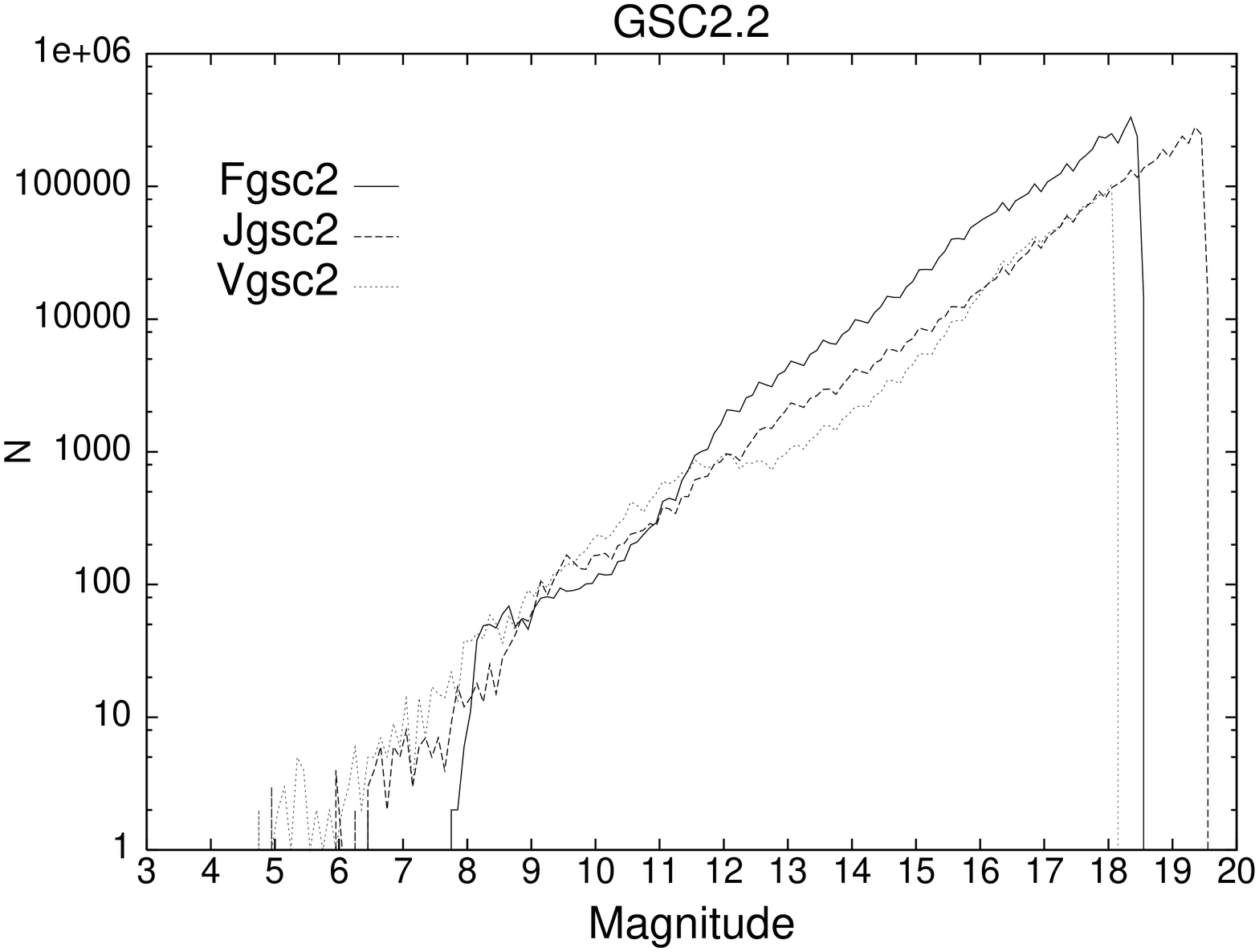,clip=,height=5cm,angle=-90} & \psfig{figure=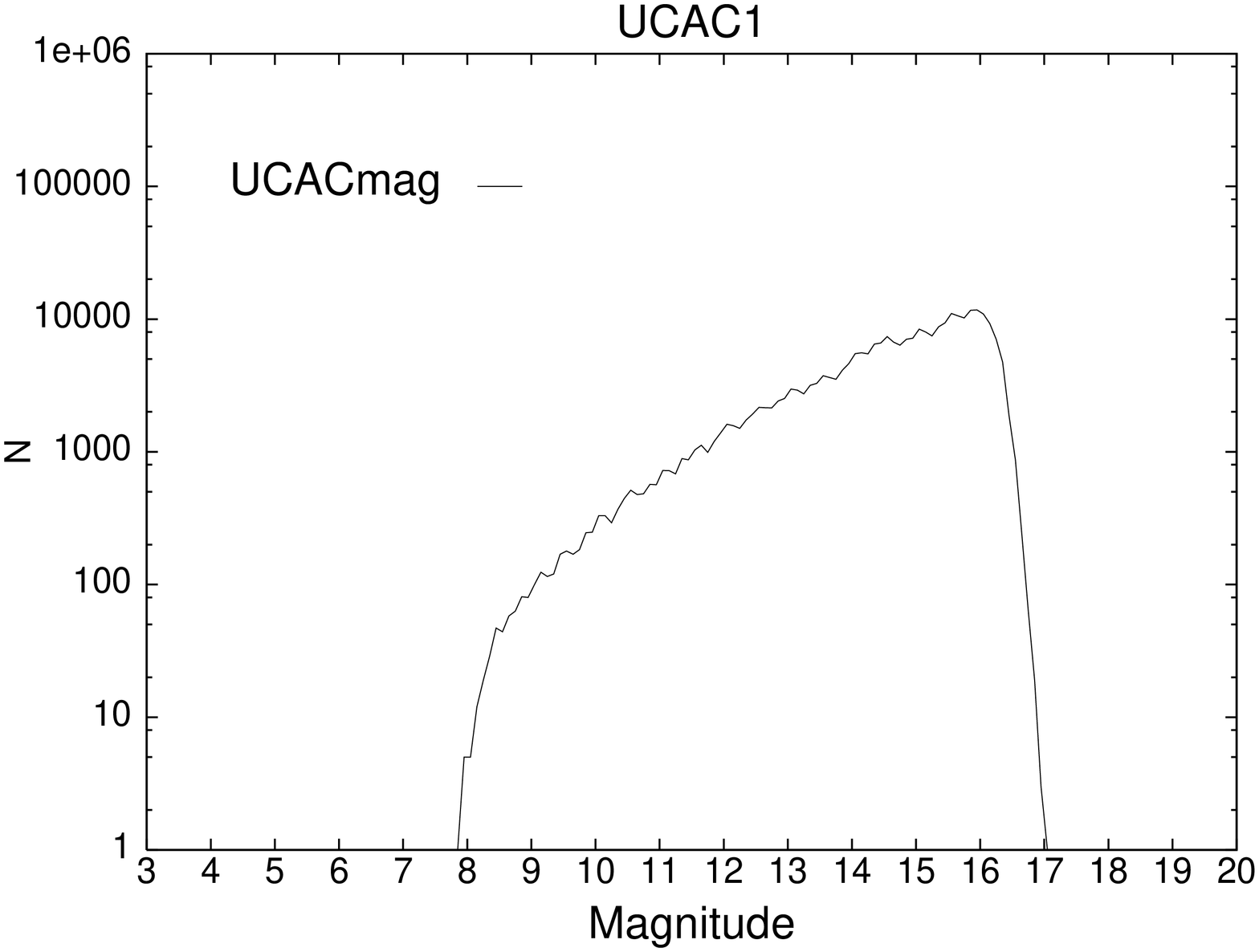,clip=,height=5cm,angle=-90} \\
   \end{tabular}
\end{center}
\caption[example] 
   { \label{fig:compl}	 
Completeness diagrams for four major surveys covering the LMC. Each plot gives the number of sources per magnitude bin.
The bin size is 0.1 magnitude. 
Note that: 2MASS observations (upper right)
are deeper than DENIS observations (upper left) in the $K_\textrm{s}$ band; the GSC2.2 magnitudes (bottom left) show a sharp cut--off ; 
the UCAC1 magnitudes (bottom right) are only indicative, since the UCAC1 is not a photometric catalogue and furthermore this is only
a preliminary catalogue, which means some improvements are expected in future releases.
   }
\end{figure*}

The Magellanic Clouds (MCs) are among the best suitable places to study the stellar evolution 
outside the Milky Way, because of their proximity and their various stellar populations.
Near--infrared surveys provide useful data for this kind of 
study because of their insensitivity to interstellar reddening.
The Magellanic Clouds have been recently fully observed by two major infrared 
surveys: the DEep Near--Infrared Survey of the Southern Sky~--~DENIS (Epchtein et al.\ 1997)  and
 the Two Micron All Sky Survey -- 2MASS (Skrutskie et al.\ 1997).
A Near--Infrared Point Source Catalogue towards the Magellanic Clouds, based on DENIS data,
has been published (Cioni et al.\ 2000a; DCMC). 
The part of this catalogue devoted to the Large Magellanic Cloud (LMC) 
covers an area of $19.87 \times 16$ square degrees
centered on ($5^\mathrm{h}27^\mathrm{m}20^\mathrm{s}$,~$-69^{\circ}00\arcmin00\arcsec$).
To compile this catalogue, the objects were required to be detected in at least 
two of the three DENIS bands $I (\mathrm{Gunn-}i, 0.79\mu \mathrm{m})$, $J (1.22\mu \mathrm{m})$, $K_\textrm{s} (2.15\mu \mathrm{m})$.
The 2MASS project observed the whole Magellanic Clouds in three photometric bands:
$J (1.23\mu \mathrm{m})$, $H (1.63\mu \mathrm{m})$ and $K_{\mathrm{s}} (2.15\mu \textrm{m})$.
For this work we used only the data available from 
the 2nd Incremental Release 
PSC\footnote{\tt http://www.ipac.caltech.edu/2mass/}, which do not cover two
rectangular regions crossing the bar of the Large Magellanic Cloud and some cross--like gaps around bright stars.
{ ($4^{\mathrm{h}}00^{\mathrm{m}}00^{\mathrm{s}}$ $<$ R.A. $<$ $7^{\mathrm{h}}00^{\mathrm{m}}00^{\mathrm{s}}$; $-78^{\circ}01\arcmin37\arcsec$ $<$ Dec. $<$ $-60^{\circ}48\arcmin00\arcsec$)},    

The number of sources from both surveys are recorded in Table~\ref{tab:nbsource}. 
Because of different sensitivity limits, DENIS sources detected only in the $I$ and $J$ bands are often detected
in $H$ and $K_\textrm{s}$ by 2MASS.
2MASS observations are more than one magnitude deeper than DENIS in the $K_\textrm{s}$ channel 
(due to a better thermalization), while they
are roughly equivalent in the $J$ channel (Fig.~\ref{fig:compl}). Thus it appeared very
interesting to cross--match the two catalogues to complete the spectral range 
of the DCMC $IJ$--sources with the $H$ and $K_\textrm{s}$ bands coming
from 2MASS, though observations are not simultaneous. 

More generally, cross--matching catalogues is highly relevant for completing the spectral or spatial coverage
when there are missing or unpublished data. It is also a powerful tool to cross--validate the catalogues and search 
for discrepancies. 

Cross--matching infrared (IR) with optical catalogues, such as DCMC/2MASS with the Guide~Star~Catalog~II (GSC2.2), 
helps on producing new colour--magnitude and colour--colour
diagrams, thus offering multispectral views of the LMC. In the cross--matching procedure we also included 
the proper motions from the USNO CCD Astrograph Catalogue (UCAC1), in order to discriminate MC members from foreground stars.
The resulting MC2
catalogue provides an unprecedented basis for the study of stellar populations in the
Magellanic Clouds and for further cross--identifications with catalogues at other wavelengths. 

{ Section 2 gives an overview of each survey towards the LMC. Section 3 deals with the strategy developed 
to cross--match the infrared DENIS and 2MASS catalogues. 
Following in Sect. 4 is a comparison of the DENIS and 2MASS photometric systems. 
In Sect. 5 we add the optical GSC2.2 and UCAC1 catalogue to the cross--matching procedure.
In Sect. 6 we present a few multispectral views of the stellar populations of the Clouds, based on the MC2 data.
}
\begin{table} [h]   
\small
\begin{center}
	\caption{Number of sources as a function of detected wavebands in the DCMC and 2MASS catalogues. 
        }
	\label{tab:nbsource}
	\begin{minipage}[r]{\linewidth}
		\centering
			\begin{tabular}{rr|rr}
				\multicolumn{4}{c}{\bf LMC}  \\
				\hline
				\multicolumn{2}{c|}{DCMC} & \multicolumn{2}{c}{2MASS}\\
				\hline
				$IJK_\textrm{s}$ & 297,031 & $JHK_\textrm{s}$ & 1,996,382 \\
				$IJ$ & 1,151,789 & $JK_\textrm{s}$ & 66 \\
				$IK_\textrm{s}$ & 8,724 & $JH$ &  - \\
				$JK_\textrm{s}$ & 1,897 & $HK_\textrm{s}$ & 4 \\
				 &  & $J$ & 11 \\
				 &  & $H$ & - \\
	 			 &  & $K_\textrm{s}$ & 23 \\
				 &  & {\it Saturated} & 259 \\
				\hline 
				Total & 1,459,441 & Total & 1,996,745 \\
				\hline 
			\end{tabular}
	\end{minipage}
\end{center}
\end{table}

\section{Data Overview} 

\begin{figure*}
\begin{center}
	\begin{tabular}{cc}
		DCMC & 2MASS\\
		\psfig{figure=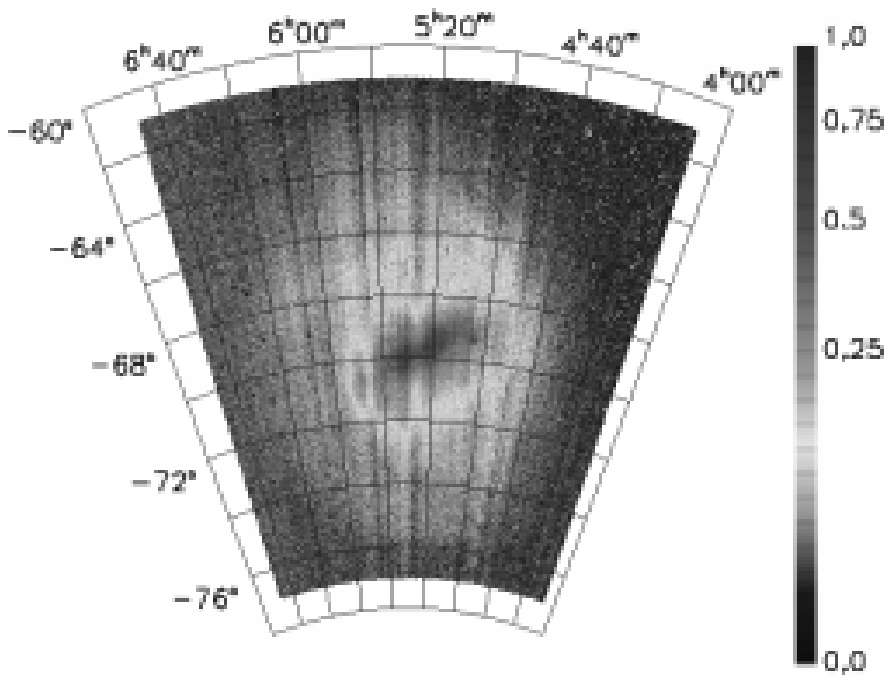,clip=,height=5cm} & \psfig{figure=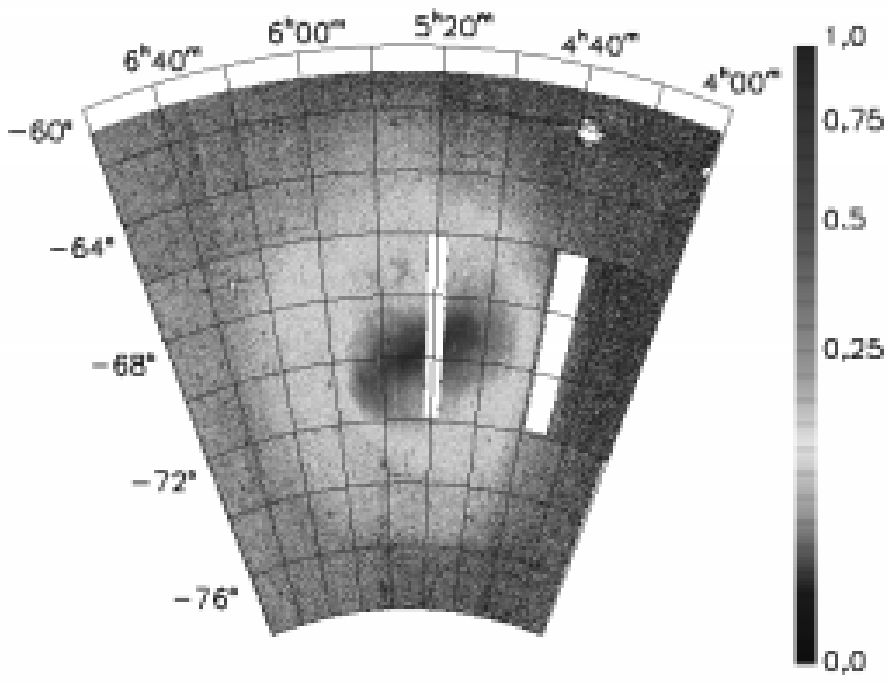,clip=,height=5cm} \\
		GSC2.2 & UCAC1\\
		\psfig{figure=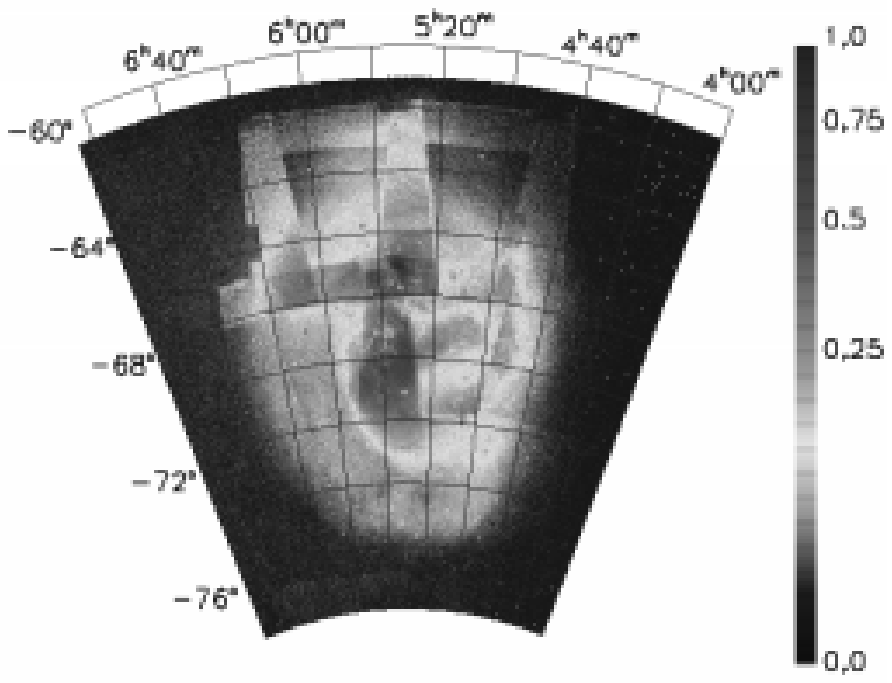,clip=,height=5cm} & \psfig{figure=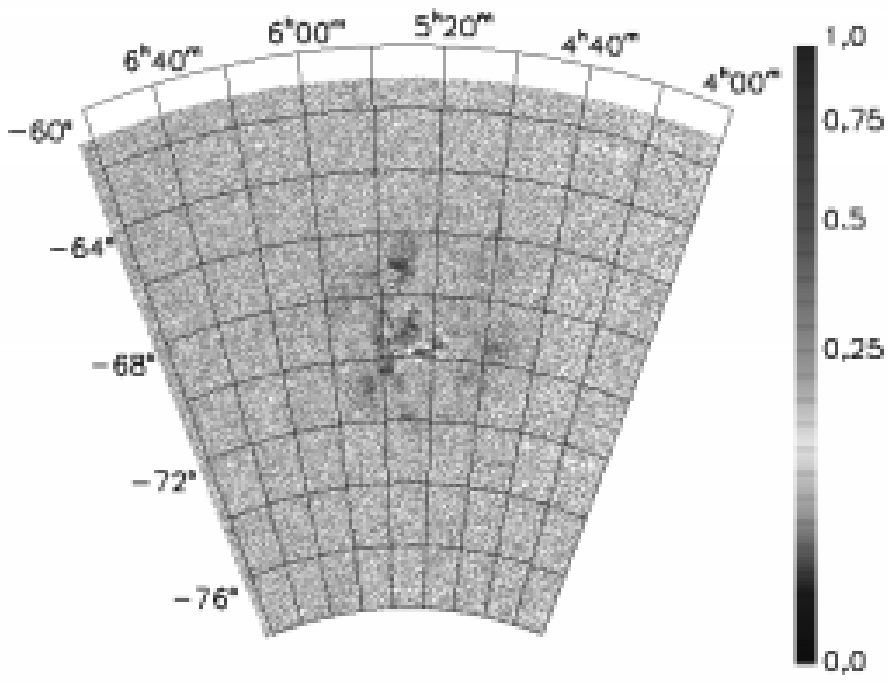,clip=,height=5cm} \\
	\end{tabular}
\end{center}
\caption[example]
   { \label{fig:map}	  
{Density maps of the LMC: DCMC (upper left), 2MASS (upper right), GSC2.2 (bottom left) and
 UCAC1 (bottom right) sources. The pixel size is $\sim 2.8^{\prime} \times 2.7^{\prime}$.
 Each map has been normalized to unity, but note that the highest pixel value reaches 208 for the DCMC, 186 for 2MASS,
 656 for the GSC2.2 and 18 for the UCAC1. }
   } 
\end{figure*} 

The present work is based on public data from 2MASS, as described in Sect.~1, and on
DCMC data obtained from a local copy of the catalogue that includes the missing strips
of the first release (the second release is currently under process). The GSC2.2 and the UCAC1 catalogues
are now also publicly available.
For each catalogue, a density map is produced (see Fig.~\ref{fig:map}).
The higher concentration of sources in the upper left part of each plot is in the direction of the Galactic center.

Inhomogeneities in the magnitude limit 
of some DENIS strips are visible on the DCMC map. 
White areas in the 2MASS map denote missing data due to observations not yet released. The circular gap on the upper 
right corner corresponds to the semi--regular pulsating star HD 29712, which is also 
the second brightest star in the sky in
$K_\textrm{s}$ (after Betelgeuse) with a magnitude lower than -4.

The GSC2.2\footnote{\tt http://www-gsss.stsci.edu/gsc/gsc2/GSC2home.htm} catalogue 
provides, in the direction of the LMC, 
{ ($3^{\mathrm{h}}59^{\mathrm{m}}30^{\mathrm{s}}$ $<$ R.A. $<$ $7^{\mathrm{h}}04^{\mathrm{m}}30^{\mathrm{s}}$; $-78^{\circ}00\arcmin00\arcsec$ $<$ Dec. $<$ $-61^{\circ}00\arcmin00\arcsec$)},    
the $F_\textrm{gsc2}$, $J_\textrm{gsc2}$, and $V_\textrm{gsc2}$ photographic bands for 6,032,541 entries.
The $V_\textrm{gsc2}$ magnitudes are from short exposure $V$ plates used to deal with dense regions of the sky.
{The photographic magnitudes given in the GSC2.2 are in the natural systems of the photographic plates
(emulsion/filter: IIIaF+OG590, IIIaJ+GG395, IIaD+W12).
    The photometric calibration is based on a Chebyshev polynomial fit to sequence stars from the 
    Second Guide Star Photometric Catalogue (Bucciarelli et al. 2001) and the Tycho Catalog for the bright end. 
       }  
However the three magnitudes are not always present together. 
The $J_\textrm{gsc2}$ band is missing for the 
innermost parts of the LMC, but is replaced by the $V_\textrm{gsc2}$ band. Thus, when using the $V_\textrm{gsc2}$ band
in the following colour--magnitude or colour--colour diagrams of this paper, one should keep in mind that we are dealing
with the central parts of the LMC only.
The unusual patterns on the GSC2.2 density map match the HTM (Hierarchical Triangular Mesh) 
partitioning of the data, 
which is a recursive spatial indexing scheme dividing the unit
sphere into spherical triangles.
{This is probably due to difficulties in producing the final
catalogue for the most crowded regions.}   
    
The UCAC1 catalogue\footnote{\tt http://ad.usno.navy.mil/ad/ucac/} (Zacharias et al.\ 2000) contains, in the direction of the LMC,
267,103 entries
{($3^{\mathrm{h}}59^{\mathrm{m}}30^{\mathrm{s}}$ $<$ R.A. $<$ $7^{\mathrm{h}}04^{\mathrm{m}}28^{\mathrm{s}}$; $-78^{\circ}00\arcmin00\arcsec$ $<$ Dec. $<$ $-61^{\circ}00\arcmin00\arcsec$)}.    
This is the preliminary version of an astrometric catalogue, which aims at increasing the number of 
optical sources with high positional accuracy. Proper motions are available, combining the UCAC1
with the USNO-A2.0 (Monet et al. 1998) positions for faint stars, and with older catalogues for bright stars. 
One magnitude, intermediate
between Johnson $V$ and $R$, is provided (579-642~nm). White regions on the UCAC1 density map denote 
missing digitalization frames in the center of the bar, due to difficulties in extracting sources in overcrowded
regions of the sky.

The catalogues used to build the MC2 present the following differences:
the observational strategy that influences the homogeneity of the final data, the passbands, the 
characterized stellar populations and the number of sources. 
These factors have a strong impact on the results of our cross--matching.


\section{First step: cross--matching DENIS and 2MASS}
\label{sect:cross}

\subsection{Cross--matching strategy}
\label{subsect:DCMCvs2MASS}

Before running the cross--matching programs, we organized the original data,
splitting most of the catalogues into smaller pieces.
The DENIS observational strategy had been to divide the sky in strips 
of $30^{\circ}$ in Declination (Dec.)
and $12^{\prime}$ in Right Ascension (R.A.).
To define subsamples, we split the DCMC catalogue by strip number because:
\begin{itemize}
\item our cross--matching algorithm is well adapted to data files with small extension in RA;
\item the cross--matching criterion depends on the strip number 
as explained below. 
\end{itemize} 
There are 119 strip--files covering the LMC.
For each strip--file, we extracted from the 2MASS catalogue all the point sources 
belonging to the same region of the sky.
Matches between both catalogues are found by specifying a position search box 
of a few arcseconds, and comparing
the coordinates of entries in both catalogues. 
The cross--matching program is executed for each strip, 
starting, each time, with two input 
files, one from DCMC and one from 2MASS.  
Both files have been previously sorted by declination, in order to optimize 
the cross--comparison procedures. 
Details about the procedure are as follows: 
for each record of the first file (say, DCMC) we search for all possible 
cross--matches in the second file (here 2MASS). 
Among the possible cross--matches, we only keep the one with
the smallest difference in position 
as the most probable counterpart.
The actual limits imposed to the positional difference $\Delta\alpha$ and $\Delta\delta$ 
depend upon the relative astrometry of the strip (see below).

\subsubsection{Finding discrepancies in the original catalogues}
\label{subsubsect:discrep}

Cross--matching by position works very well in most cases because the astrometry
of DCMC and 2MASS is accurate enough (better than one arcsecond). 
{2MASS positions were reconstructed from the ACT reference
catalogue (Urban et al., 1998), using the Tycho astrometry.  The astrometric
reference for DCMC
positions is the USNO-A2.0 catalogue (Monet et al. 1998).
The astrometric solution is global for a strip,
minimizing possible inaccuracies of the USNO-A2.0 catalogue
in the most crowded regions. }  

Consequently the match distance is smaller 
than $0.5^{\prime\prime}$ for the great majority of the stars. There is in principle no risk of
confusion at such a small scale. While this is true in general, 
in practice the cross--matching exercise has proven to be a 
powerful tool to detect subsets of data which deviate from 
the perfect situation, and primarily areas suffering from problems
in the astrometric or photometric calibration.

In some cases, field distortions in the DCMC affect the quality of the astrometry. To detect and quantify them, 
we proceeded strip by strip. We kept only well confirmed DCMC sources: $10.5 \leq I \leq 16.5$ and flags in the $I$
band equal to zero. We ran a cross--matching program based only on distances, with a searching box
that goes up to 30$^{\prime\prime}$. Between all the possible associations found, we kept only the 
association with $|J_{\textrm{DCMC}} - J_{\textrm{2MASS}}| \leq 0.5$. The selection is done on magnitude
because in case of field distortions, small distances are not reliable enough a criterion.

The relative shifts in R.A. and Dec. are a function of the pixel coordinates of the camera. We found 11 strips 
affected by field distortions at a level larger than $2^{\prime\prime}$.
We also searched for systematic shifts $\delta J$ and $\delta K_\textrm{s}$ between DCMC and 2MASS magnitudes.
Mean shifts have been computed for each strip.
The diagrams corresponding to the positional and magnitude shifts are all available, strip by strip,
on the MC2 web site\footnote{\tt http://vizier.u-strasbg.fr/MC2/}.

Such astrometric and magnitude shifts depend on the particular strip 
and had to be taken into account in the DENIS versus 2MASS cross--matching. 
Strategies for coping with them have been implemented,
to allow a proper strip by strip cross--matching of both catalogues. 
We took advantage of the $J$ and $K_\textrm{s}$ common magnitudes of the two surveys.
A potential cross--matched source is thus validated not only on a positional criterion, but also
on magnitude criteria.

\subsubsection{ Defining a positional searching box}
\label{subsubsect:coord} 
{    Shifts in R.A. ($\delta\alpha$) and Dec. ($\delta\delta$) being mainly a function
    of the pixel coordinates, they do vary inside one image, but are nearly the same
    for all the images of the strip.}
     So it is better to use the statistics of the whole strip instead
 of one single image. 
         Thus we can define a specific position search box for the strip. The size of 
the box will take into account the shifts
          in R.A. and Dec. found for this strip number.
          The default size of the searching box when there are no shifts is $3^{
\prime \prime}$.
          So we have now an enlarged and asymmetric searching box:

       \[        \frac{\delta \alpha_\textrm{min}^{\prime\prime} - 3^{\prime\prime}}{ \cos \delta} < \alpha_\textrm{DCMC} - \alpha_\textrm{2MASS} <  \frac{\delta        \alpha
_\textrm{max}^{\prime\prime}  +3^{\prime\prime}}{\cos \delta}
          \]           
          \[              \delta \delta_\textrm{min}^{\prime\prime} - 3^{\prime\prime}       < \delta_\textrm{DCMC} - \delta_\textrm{2MASS} < \delta  \delta_\textrm{max}^{\prime\prime}    +3^{\prime\prime}     \quad  ,
        \]
        
        where $\delta \alpha_\textrm{min} $,  $\delta  \alpha_\textrm{max} $,   $\delta \delta_\textrm{min} $,  $\delta  \delta_\textrm{max}  $
        are the minimum and maximum shifts in R.A. and Dec. (arcseconds).
Note that the searching box has a complex shape, since $\delta \alpha_\textrm{min} \neq \delta  \alpha_\textrm{max}$ and
 $\delta \delta_\textrm{min} \neq \delta  \delta_\textrm{max}$.
 This box is used to optimise the probability to find the correct cross--matching, even in distorted images.

\subsubsection{Selection on magnitudes}

     Between all the possible associations found in Sect.~\ref{subsubsect:coord}, we must keep the best one. 
     We have seen that keeping the association with the smallest distance is no more a
     reliable criterion because of field distortions. So we have to check the compatibility in
     magnitude for each association, after applying on the strip data the associated mean magnitude shifts 
     $<\delta J>$ and $<\delta K_\textrm{s}>$ computed in Sect. \ref{subsubsect:discrep} above.

\begin{itemize}
\item         If $K_\textrm{s}$ is not detected in the catalogues, the selection is done on $J$. The following 
relation has to be true to keep the association:
\begin{eqnarray*}
        |\delta J - <\delta J>|  \leq w  \times \sqrt {\sigma_
{J_\textrm{\scriptsize DCMC}}^{2} +    \sigma_{J_\textrm{\scriptsize  2MASS}}^{2}}    \quad,
\end{eqnarray*}         
where $w=2$ is a weight, and $\sigma_{J_\textrm{\scriptsize DCMC}}$ and
 $\sigma_{J_\textrm{\scriptsize        2MASS}}$ are the relative
photometric uncertainties as quoted in both catalogues. Relative uncertainties
are in general very small for bright stars, less than 0.01 mag. However, uncertainties on the absolute 
calibration are much larger: about 0.1 mag for the DCMC. If we apply abruptly
the above criterion, we will lose many cross--identifications for the stars with
 small relative uncertainties.
We thus need to refine the selection criterion and consider two cases:

      \[ \textrm{if} \qquad w \times \sqrt {\sigma_{J_\textrm{\scriptsize 
DCMC}}^{2} + \sigma_{J_\textrm{\scriptsize           2MASS}}^{2}} \leq 
              \Delta      J \qquad \textrm{then} \]
    \[  |\delta J - <\delta J>| \leq \Delta J \quad  \]  
     \[ \textrm{else if} \qquad w \times \sqrt {\sigma_{J_\textrm{\scriptsize 
DCMC}}^{2} + \sigma_{J_\textrm{\scriptsize               2MASS}}^{2}} >
        \Delta J            \]  \[   \textrm{then}  \qquad   |\delta J -
 <\delta J>|  \leq w  \times \sqrt      {\sigma_{J_\textrm{\scriptsize DCMC}}^{2} +      \sigma_{J_\textrm{\scriptsize  2MASS}}^{2}} 
\]

where  $\Delta      J=0.45$ is the estimated full-width at half-maximum 
of the $\delta \textrm
{J}$ distribution.  

  \item        If $J$ is not detected in the catalogues, the selection is done 
  on $K_\textrm{s}$ as above but this time we have  $\Delta  K_\textrm{s} = 0.60$.
 \item       If $J$ and $K_\textrm{s}$ are detected in both catalogues, the 
 selection is done on $J$ and then on $J$--$K_\textrm{s}$. 
   \item        {  If neither $J$ nor $K_\textrm{s}$ are detected}  in the 
catalogues, the association is lost. 
         
\end{itemize}

 Applying these criteria, if there are still more than one possible association for one DCMC source, then
     we keep the association with the smallest $\delta J$ or $\delta K_\textrm{s}$.

\begin{figure}
\begin{center}
   \begin{tabular}{c}
 	  \psfig{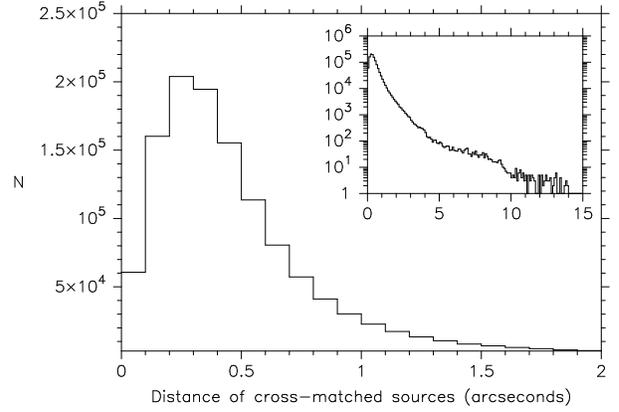} \\
   \end{tabular}
\end{center}
\caption[example] 
   { \label{fig:distXID}	 
Results of the cross--matching between DCMC and 2MASS. Number of objects as a function
of the distance of the cross--matched point sources.
The bin size is $0.1^{\prime\prime}$.
   }
\end{figure}

\begin{figure*}
\begin{center}
	\begin{tabular}{cc}
		\psfig{figure=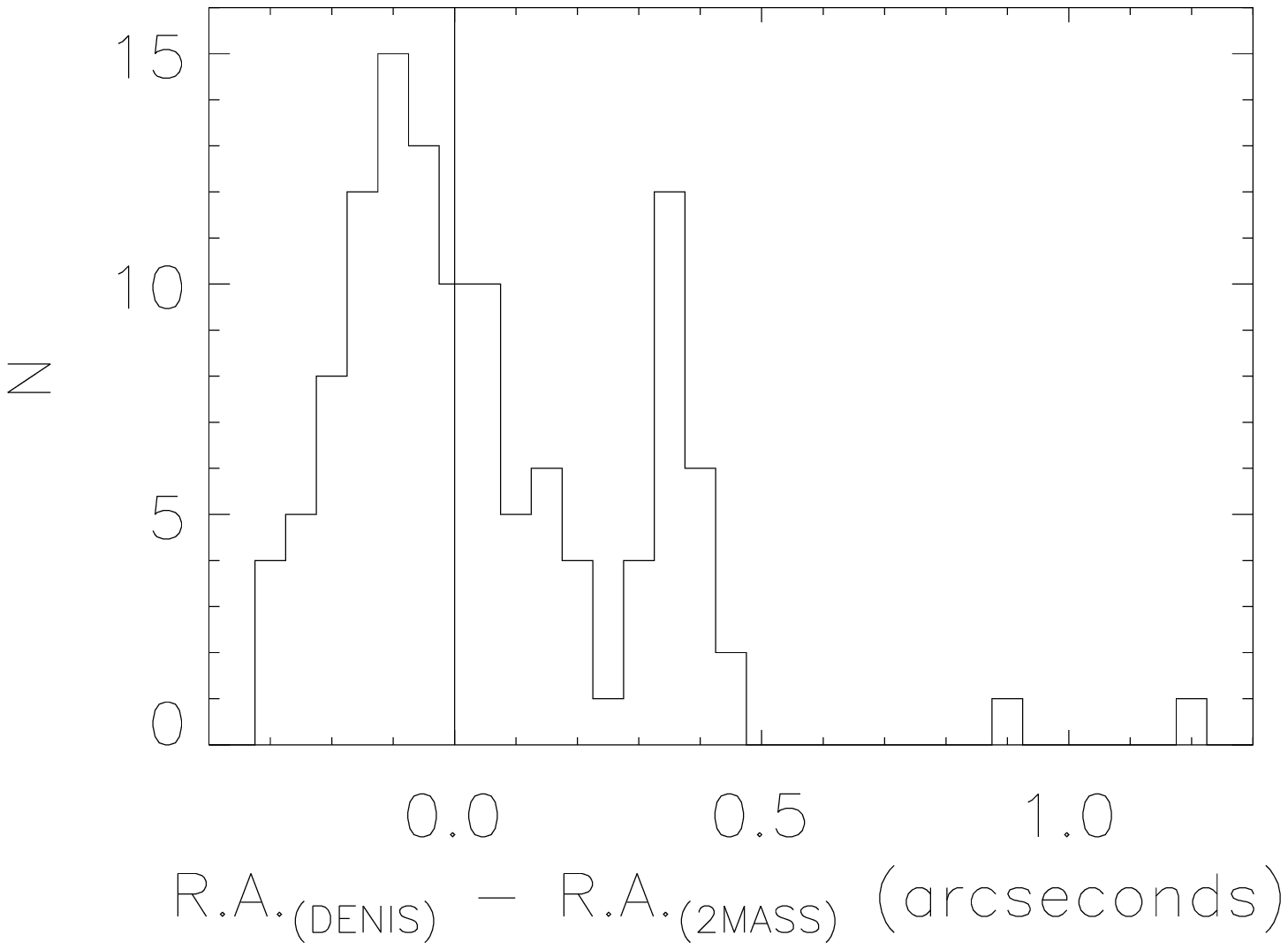,width=0.9\columnwidth} & \psfig{figure=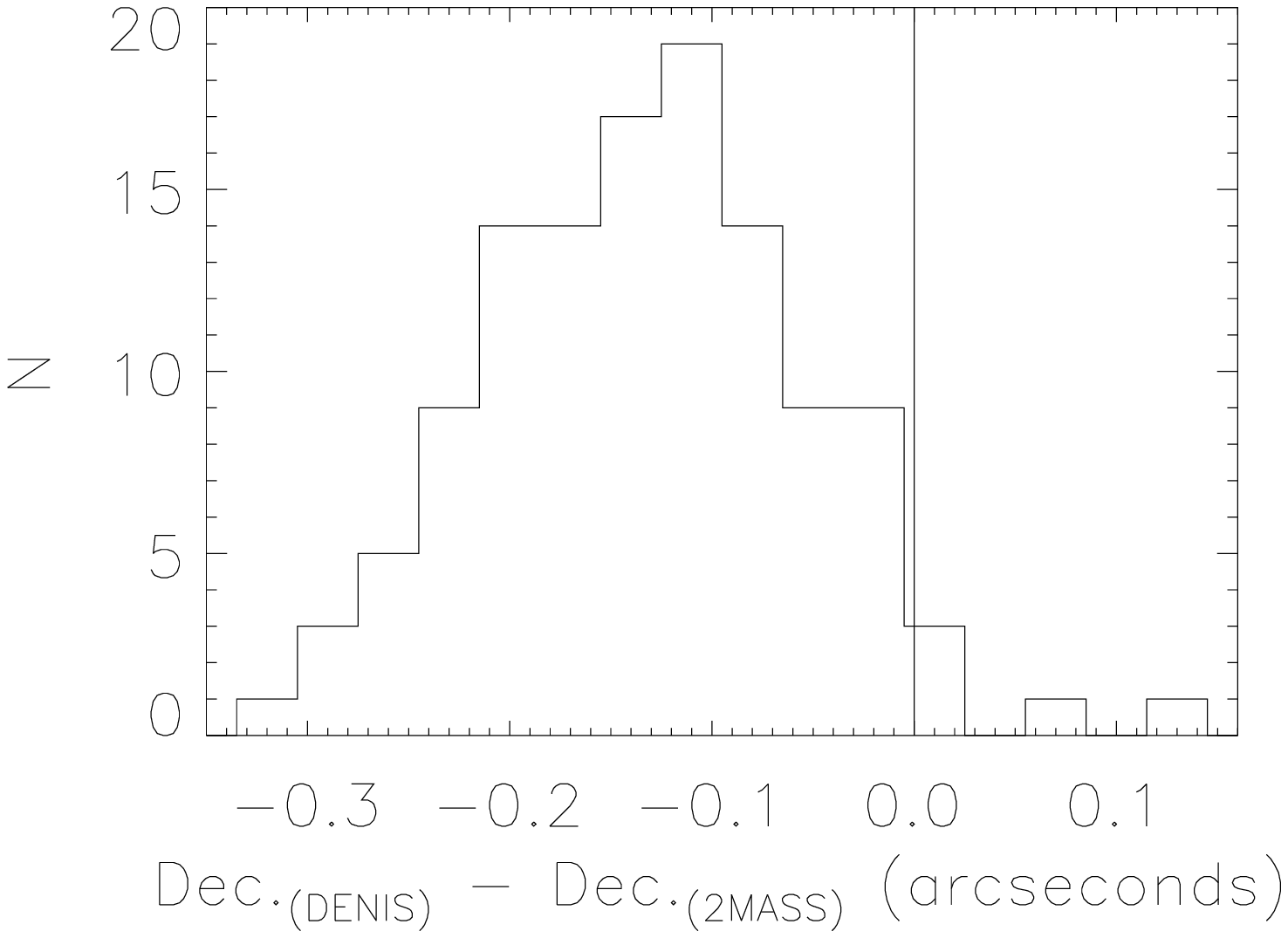,width=0.9\columnwidth}\\
	\end{tabular}
\end{center}
\caption[example] 
   { \label{fig:shiftRADEC}	  
Histograms of the shifts between DCMC and 2MASS $\alpha$ and $\delta$ (in arcseconds) for the 119 strips covering the LMC.
The bin size is $0.05^{\prime\prime}$ for R.A. and $0.03^{\prime\prime}$ for Dec.
   } 
\end{figure*}

More details about this cross--matching step, as well as
the cross--matching criteria used can be found in Delmotte et al. (2001).
{Nearly 80\% of the LMC strips 
have a match rate better than 90\%. The strips with a match rate 
smaller than 80\% correspond   to the gaps in the 2MASS data.}
{We checked the distance distribution of the matches, by wether they were done in $J$, or $K_\textrm{s}$
or both. There seems to be no relation between the magnitude criterion applied and
the distance of the cross--matched source.
Figure~\ref{fig:distXID} shows the results of the cross--matching between DCMC and 2MASS, whatever the magnitude criterion was.
{The mean positional offset between matches is $0.52^{\prime\prime}$ and the modal offset is $0.25^{\prime\prime}$}.}
{ 
Figure~\ref{fig:shiftRADEC} 
displays the histograms of the shifts between DCMC and 2MASS in R.A. and Dec. (in arcseconds) for the 119 strips covering the LMC.}
{To check the results, we also compared the distribution of the close matches ($\leq 2^{\prime\prime}$) and far matches ($\geq 4^{\prime\prime}$)
in both the ($J$--$K_\textrm{s}$, $K_\textrm{s}$) colour--magnitude diagram and (R.A., Dec.) plane. Far matches do not show any strange physical behavior
and are, as expected, distributed along lines associated with the borders of the strips suffering from field distortions, and also in the center of the Cloud
where the density is higher.}

\section{Comparing DENIS and 2MASS photometric systems}

We considered the mean linear relation between DCMC and 2MASS magnitudes, 
restricting to the range [10, 14] in $J$
and [8, 12] in $K_\textrm{s}$, avoiding the saturated bright stars as well as the faintest ones.

\begin{figure*}
\begin{center}
	\begin{tabular}{cc}
		\psfig{figure=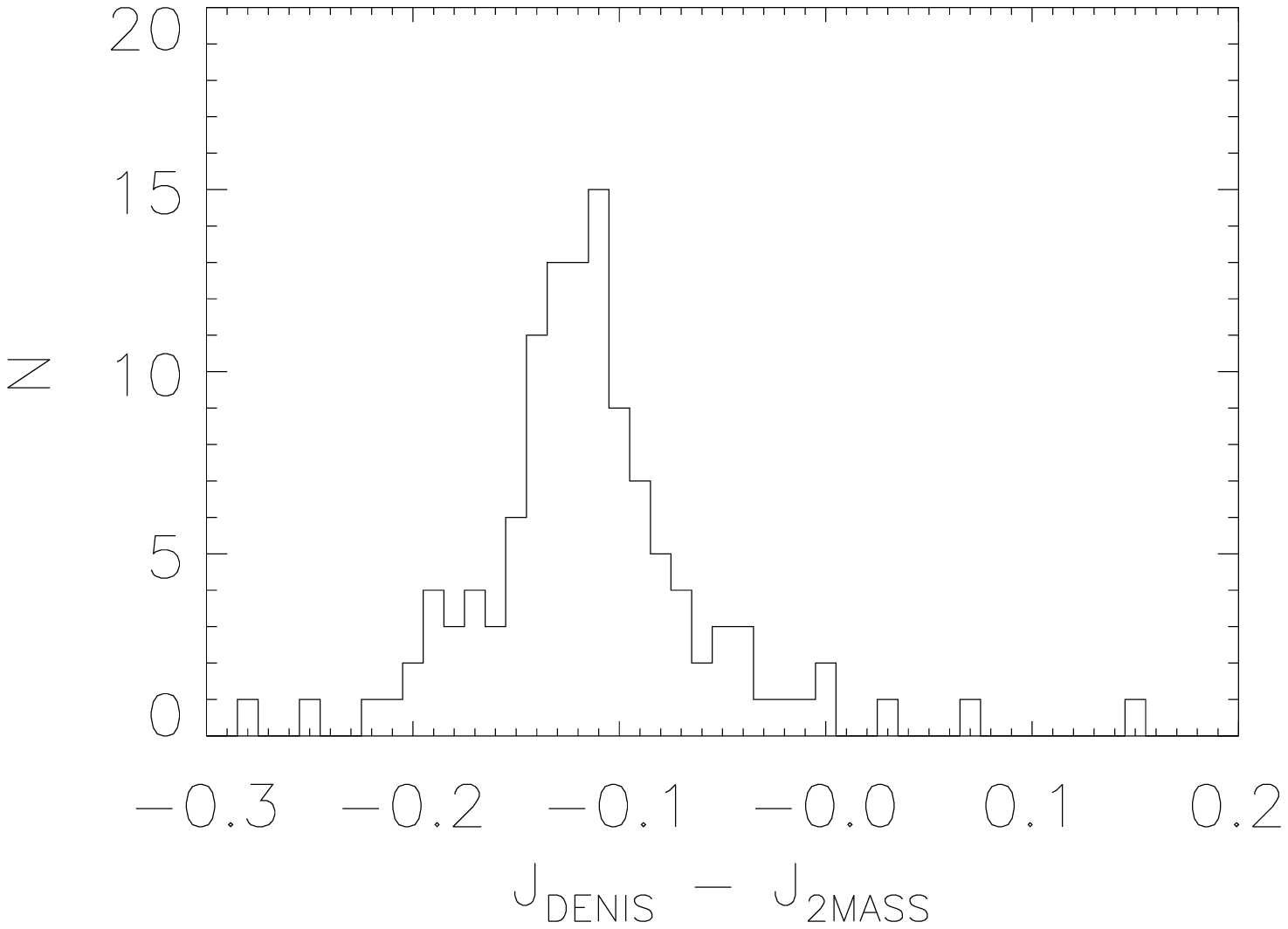,width=0.9\columnwidth} & \psfig{figure=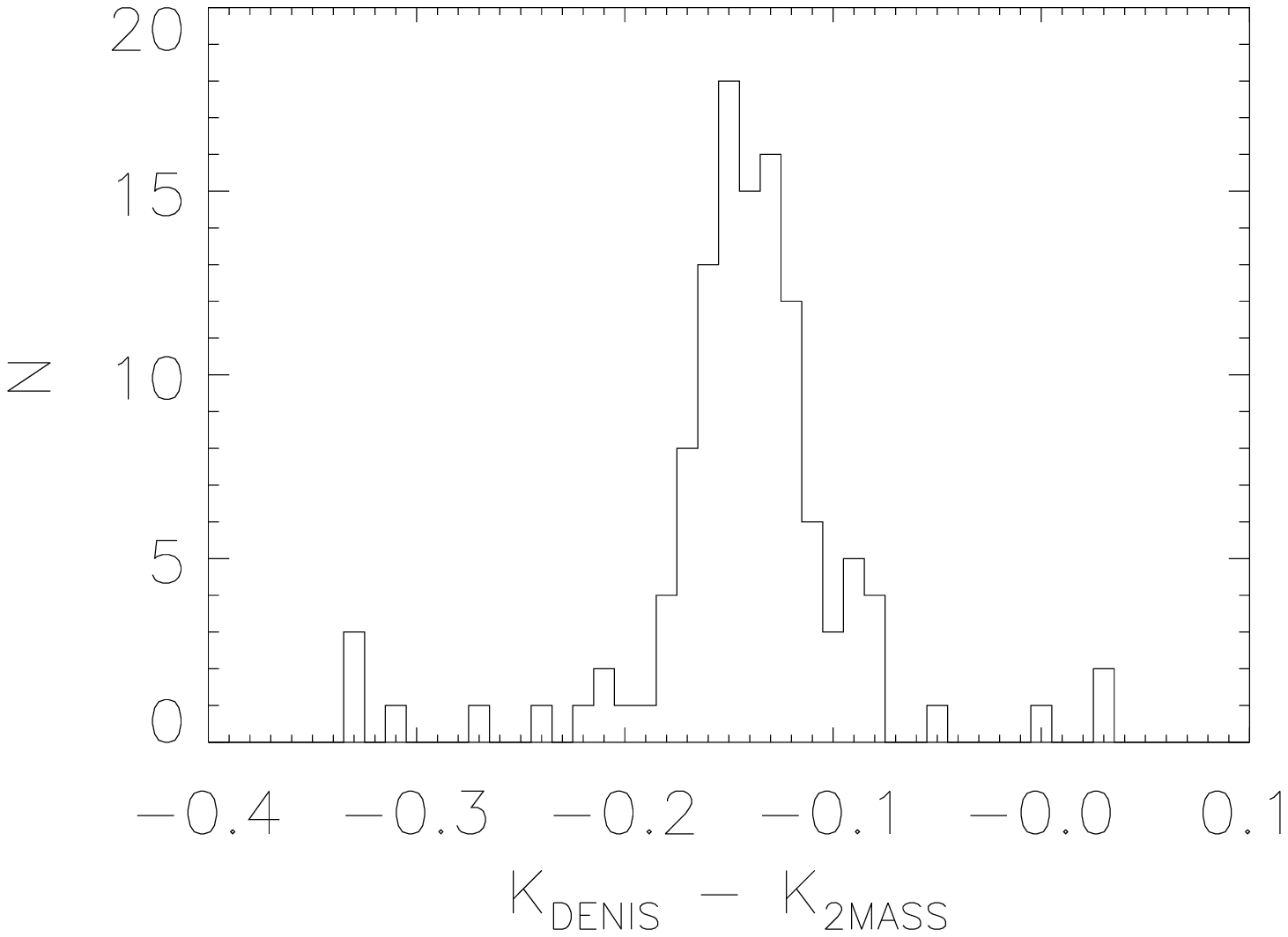,width=0.9\columnwidth}\\
	\end{tabular}
\end{center}
\caption[example] 
   { \label{fig:histodiff}	  
Histograms of the shifts between DCMC and 2MASS $J$ and $K_\textrm{s}$ magnitudes for the 119 strips covering the LMC.
 The bin size is $0.01^{\prime\prime}$.
  } 
\end{figure*}

We find a systematic shift of the absolute calibration between 
the two catalogues.
For each strip, we calculated the median 
of $\delta J$ and $\delta K_\textrm{s}$. 
Figure~\ref{fig:histodiff} shows the histograms
of the shifts found for the 119 LMC strips.

The derived mean relations between the two systems are
as follow:

$$    J_{\mathrm{DCMC}} = J_{\mathrm{2MASS}} - (0.11 \pm 0.06)  $$
$$    K_{\mathrm{sDCMC}} = K_{\mathrm{s2MASS}} - (0.14 \pm 0.05)  $$

These relations have been computed in the case of the DCMC catalogue and may not be valid for the
whole DENIS survey.
This is quite different from the relations proposed by Carpenter (2001),
based on a limited preliminary sample of a few DENIS sources.
Groenewegen (2000) did the comparison for a few hundred Cepheids towards the Clouds and 
found no significant difference between 2MASS and DENIS $J$, but found a 0.2 magnitude
shift for the $K_{\mathrm{s}}$ band data.
{This larger shift compared to the one we find is probably due to the use of only
variable stars (i.e. Cepheids).}

\section{Second step: adding GSC2.2 and UCAC1}
\begin{figure}
\begin{center}
   \begin{tabular}{c}
 	  \psfig{figure=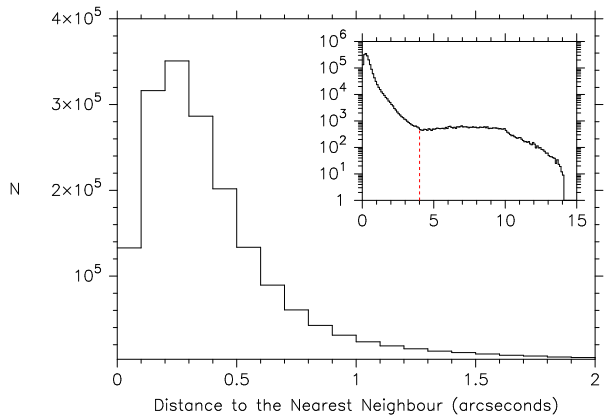,clip=,width=0.9\columnwidth,angle=-90} \\
   \end{tabular}
\end{center}
\caption[example] 
   { \label{fig:gsc2dist}	 
Results of the cross--matching between the GSC2.2 and the 2MASS catalogues. Number of objects as a function
of the distance to the nearest neighbour.
The bin size is 0.1$^{\prime\prime}$.
   }
\end{figure}

The GSC2.2 has also been divided in strips, to be cross--matched
with 2MASS. These strips are no more related to the DENIS strips, but to strips with much smaller extension 
($6^{\prime}$) in order to optimize the time needed to cross--match the catalogues.
{We chose a searching box of $10^{\prime\prime}$. It was not possible to compare the GSC2.2 and 2MASS magnitudes 
in the same way as for the DCMC vs 2MASS cross--matching (as in Sect. 3.1.3.) because there
is no common magnitude between the two surveys. As a consequence, we kept
only the associations to the nearest neighbour and then we cut the resulting distribution
at a distance of $4^{\prime\prime}$ (Fig.~\ref{fig:gsc2dist}).}
{The mean positional offset between matches is $0.45^{\prime\prime}$ and the modal offset is $0.25^{\prime\prime}$.}

\begin{figure}
\begin{center}
   \begin{tabular}{c}
 	  \psfig{figure=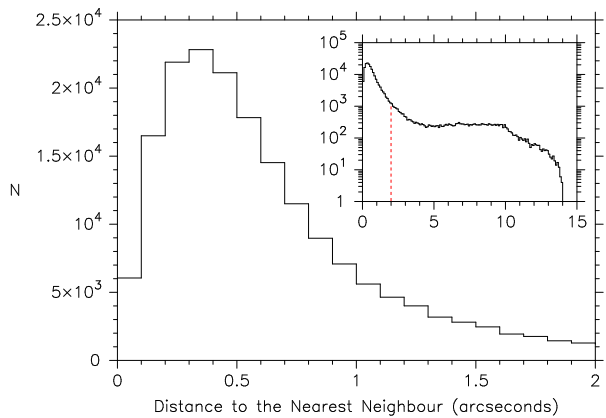,clip=,width=0.9\columnwidth,angle=-90} \\
   \end{tabular}
\end{center}
\caption[example] 
   { \label{fig:dist}	 
Results of the cross--matching between the DCMC and the GSC2.2 sources not present in 2MASS. Number of objects as a function
of the distance to the nearest neighbour. The bin size is 0.1$^{\prime\prime}$.
 }
\end{figure}

At this stage, both the DCMC and the GSC2.2 are cross--matched with the 2MASS catalogue. 
Thus the link between the DCMC and the GSC2.2 can be done using the 2MASS common point sources.
Common entries in DCMC/2MASS and GSC2.2/2MASS have been merged.
We have now six resulting files:
\begin{itemize}
\item sources present in DCMC, 2MASS and GSC2.2
\item sources present in DCMC and 2MASS only
\item sources present in GSC2.2 and 2MASS only
\item sources in 2MASS only
\item sources in the DCMC only
\item sources in the GSC2.2 only
\end{itemize}
We run another cross--matching process on the last two files to find sources which are present
in both DCMC and GSC2.2. 
For small positional differences the associations are likely true whereas at larger distances they are generally random associations,
see the distribution of the sources as a function of the distance from the nearest neighbour (Fig.~\ref{fig:dist}).
{The mean positional offset between matches is $0.61^{\prime\prime}$ and the modal offset is $0.35^{\prime\prime}$.}
  
\begin{figure*}
\begin{center}
   \begin{tabular}{ccc}
    \psfig{figure=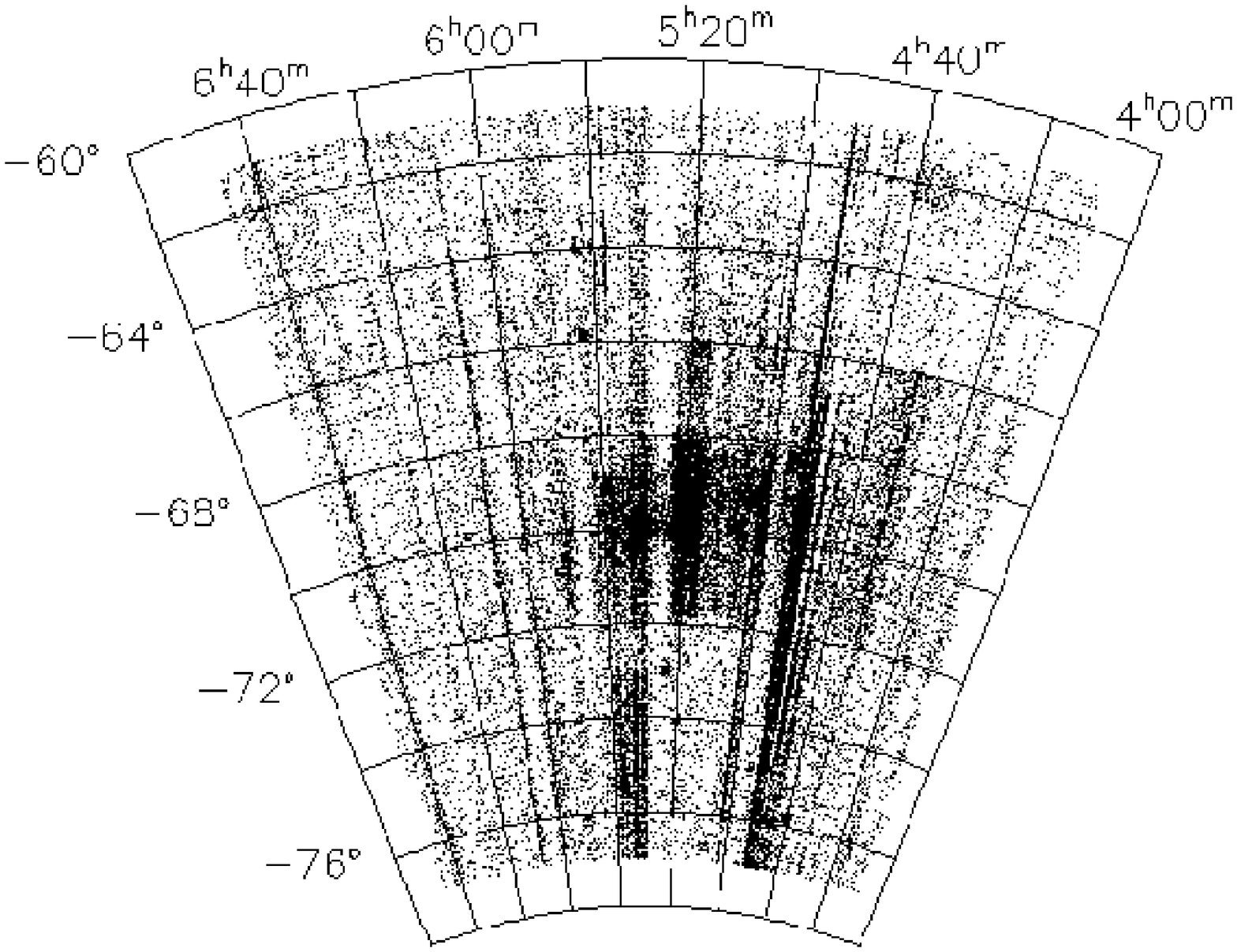,clip=,height=4.6cm} & \psfig{figure=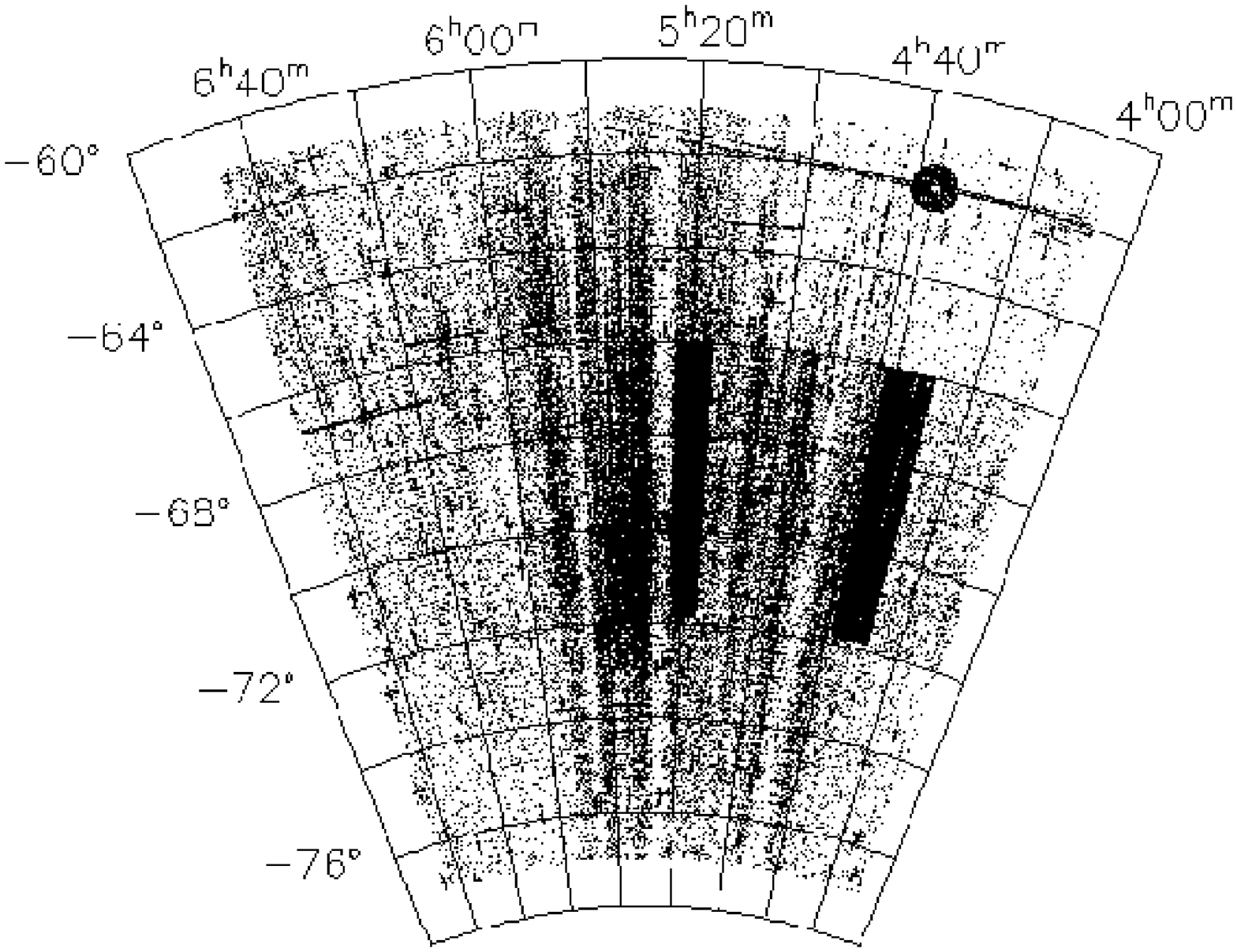,clip=,height=4.6cm}  & \psfig{figure=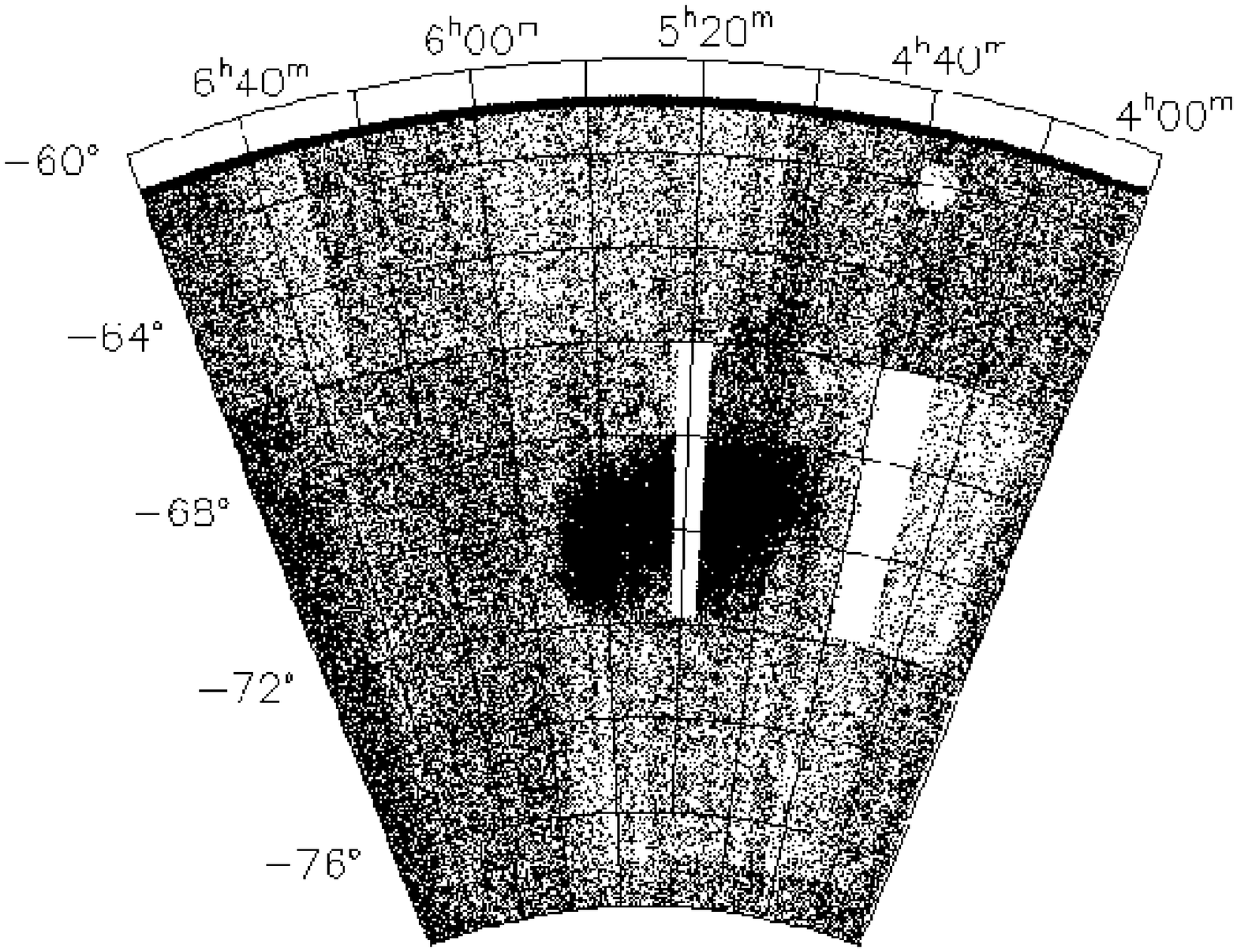,clip=,height=4.6cm} \\
   \end{tabular}
\end{center}
\caption[example] 
   { \label{fig:threeplots}
Spatial distribution of some cross--matched sources.	 
Left panel: sources of the MC2 belonging to the DCMC catalogue only. The DENIS strip structure remains visible, because some of them
suffer from field distortion all along the borders.
Middle panel: sources of the MC2 present only in 
the DCMC and the GSC2.2 catalogues. Most of them fill the gaps of the 2MASS catalogue. 
Right panel: sources of the MC2 belonging to the 2MASS catalogue only.
   }
\end{figure*}

However, because about 1\% of the DCMC sources suffer from astrometric problems, we cannot exclude
that there are some true associations for sources at larger distances. Unfortunately there is no magnitude in common
between the DCMC and the GSC2.2 catalogue and we cannot adopt the same strategy as in Sect. 3.1.3.
We decided to keep only the associations with distances smaller than $2^{\prime\prime}$, to avoid polluting
the MC2 with too many false associations. On the other hand, we are losing a few associations, located in
the poorly calibrated regions of the DCMC (see the strip like features on the left plot of Fig. \ref{fig:threeplots}).
{A comparison of the slopes of the histograms in Fig.~\ref{fig:gsc2dist} and~\ref{fig:dist} does
not seem to indicate any excess of false matches for distances between 2\arcsec~and 4\arcsec.}
Sources belonging to both DCMC and GSC2.2, but not 2MASS, are shown on the middle plot of 
Fig.~\ref{fig:threeplots}.
They correspond mainly to sources falling in the yet empty gaps of the 2MASS data.
The 2MASS scanning strategy covered the sky with tiles 6 degrees long in Dec.
and $8.5^{\prime}$ wide in R.A. These patterns remain visible on the right plot of Fig.~\ref{fig:threeplots}, showing
the spatial distribution of the MC2 sources belonging to the 2MASS catalogue only, thus denoting different sensitivity limits.

{Table~\ref{tab:MC2} summarizes the results obtained so far during the process to build up the MC2, which contains more than 6 million
sources for the LMC.
The optical/IR database contains 1,968,360 sources: IR from 2MASS (629,212), DCMC (177,414), or both
(1,161,734). Among the remaining sources, 4 million of them are only detected in the GSC2.2. 
It is of great astronomical interest to get as many wavelengths as possible for each star, but this should not lead
to disregard sources detected only with one survey and not with the other ones. Keeping non--associations
in the MC2 helps keeping track of the internal discrepancies and different sensitivity limits of each catalogue.}

\begin{table} [h]   
\small
\begin{center}
\caption{Distribution of the MC2 sources, prior the inclusion of UCAC1.}
\label{tab:MC2}
	\begin{minipage}[l]{\linewidth}
	\centering
	\begin{tabular}{c|c|c|r}
		\hline
		2MASS    &  DCMC    &  GSC2.2     & Number of sources \\
		\hline
		 \resizebox{2mm}{!}{\tick}      &     \resizebox{2mm}{!}{\tick}    &    \resizebox{2mm}{!}{\tick}       &     1,161,734     \\
		  \resizebox{2mm}{!}{\tick}      &     \resizebox{2mm}{!}{\tick}    &    ~       &        54,584     \\
		  \resizebox{2mm}{!}{\tick}      &     ~    &    \resizebox{2mm}{!}{\tick}       &       629,212     \\
		  \resizebox{2mm}{!}{\tick}      &     ~    &    ~       &       151,215     \\
	 	  ~      &     \resizebox{2mm}{!}{\tick}    &    ~       &        65,709     \\
	 	  ~      &     ~    &    \resizebox{2mm}{!}{\tick}       &     4,064,181     \\
		  ~      &     \resizebox{2mm}{!}{\tick}    &    \resizebox{2mm}{!}{\tick}       &       177,414     \\
		\hline
		\multicolumn{3}{c|}{ Total }  &     6,304,049     \\
		\hline
	\end{tabular}
	\end{minipage}
\end{center}
\end{table}

The procedure to add the UCAC1 is quite different. We cross--matched the UCAC1, without splitting it in strips,
with the MC2 at its present stage.
This is possible because the UCAC1 is a small catalogue and our program is fast enough to process it in one run.
Another advantage is that the UCAC1 is automatically cross--matched with the DCMC--only sources
and the GSC2.2--only sources. 

\begin{figure}
\begin{center}
 	  \psfig{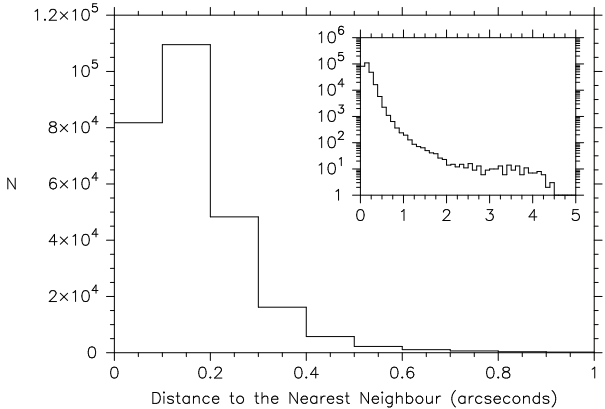}
\end{center}
\caption[example] 
   { \label{fig:ucacdist}	 
Results of the cross--matching between the UCAC1 and the MC2 catalogues. 
Histogram of distances to the nearest neighbour. The bin size is 0.1$^{\prime\prime}$.
   }
\end{figure}

\begin{figure}
\begin{center}
 	\psfig{figure=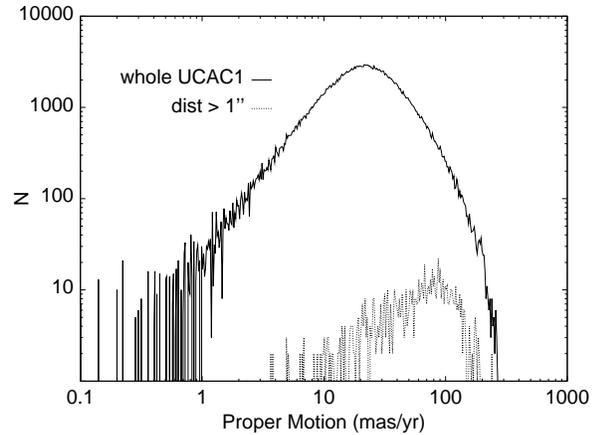,clip=,width=0.9\columnwidth,angle=-90}
\end{center}
\caption[example] 
   { \label{fig:mu}	 
Results of the cross--matching between the UCAC1 and the MC2 catalogues. 
Histogram of the proper motions. Both axes have logarithmic scale and the bin size is also logarithmic.
The dashed line refers to the cross--matched sources with distances larger than $1^{\prime\prime}$. Their distribution is
in the region of sources with large proper motions, when compared to the whole UCAC1 distribution (solid line).
   }
\end{figure}

Figure~\ref{fig:ucacdist} shows the same as Fig.~\ref{fig:gsc2dist} and Fig.~\ref{fig:dist} for UCAC1 and MC2 sources.
We decided to keep
all these associations, even the ones for sources at distances larger than $1^{\prime\prime}$, because these sources also 
display a larger proper motion compared to the average source in the catalogue (Fig.~\ref{fig:mu}).
This might as well be the cause of the large distance derived during the association process.
{The mean positional offset between matches is $0.17^{\prime\prime}$ and the modal offset is $0.15^{\prime\prime}$.}
About 42 UCAC1 sources do not have a MC2 counterpart, which means that 99.9\% of the UCAC1 catalogue is linked to 
the MC2 and 
4.2\% of the MC2 has a UCAC1 counterpart.

\section{Results: Multispectral views of the LMC} 

{This paper mainly deals with the techniques that went into the construction of the MC2. It shows
how essential a tool the cross--matching of large surveys is, to derive results on their internal
accuracy.}
The broad range of magnitudes covered by the MC2, as well as the large
number of sources involved, allow a multi--wavelength
and statistical study of the stellar populations of the Clouds.
{We present a few results concerning their location in several colour--magnitude and colour--colour diagrams, 
in order to demonstrate the usefulness of such an optical/infrared catalogue and its relevance
in the framework of the Virtual Observatory.}
Note that observations of cross--matched sources were not simultaneously performed so those following diagrams
should be considered as indicative because the colours might not represent correctly variable sources.

\subsection{The ($K_\textrm{s}$, $J$--$K_\textrm{s}$) colour--magnitude diagram}

\begin{figure*}
\begin{center}
	 \psfig{figure=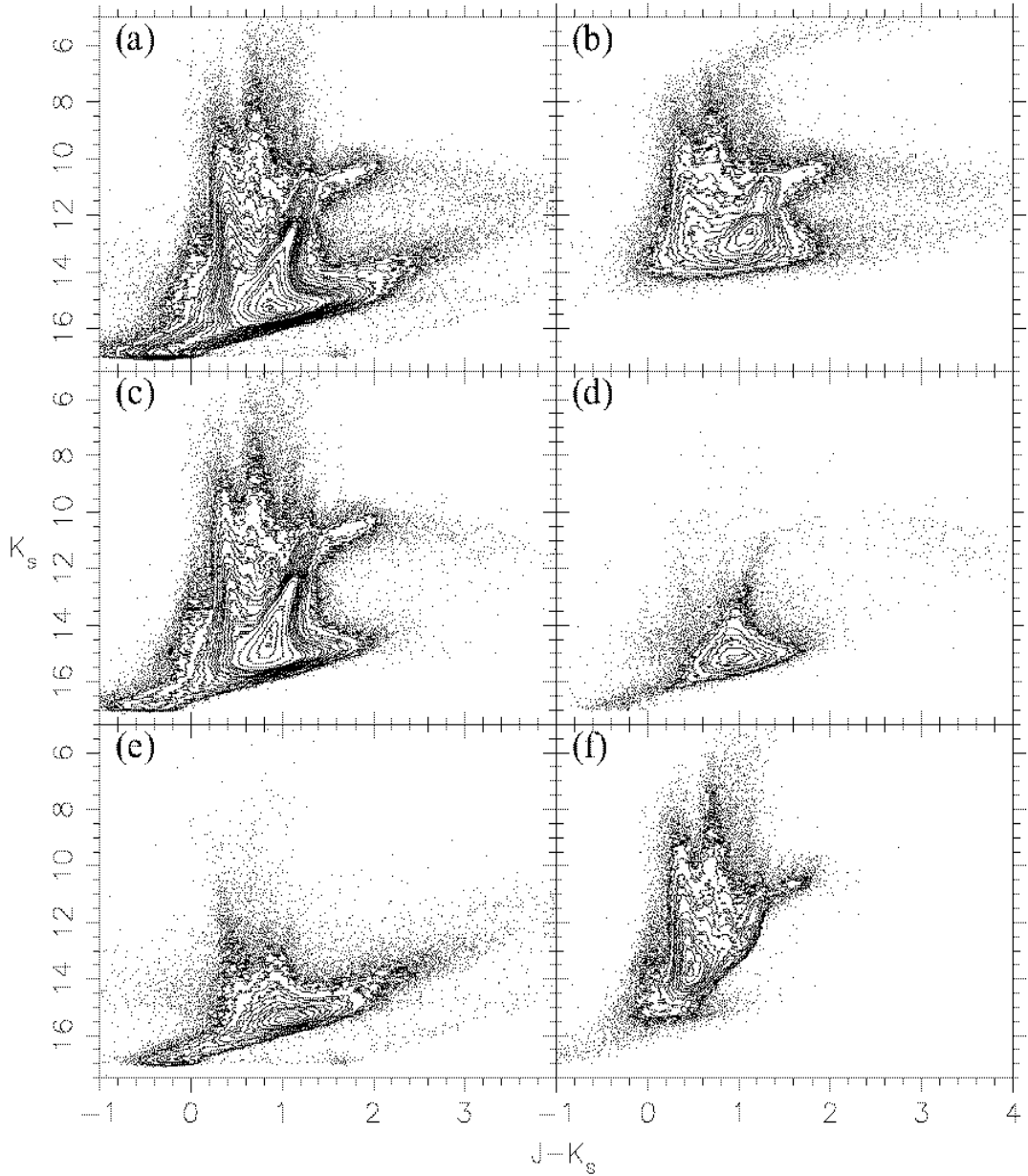,height=17cm,width=15cm}
\end{center}
\caption[example] 
   { \label{fig:comp}	
(a) ($K_\textrm{s}$, $J$--$K_\textrm{s}$) 2MASS CMD:
 1,996,448 entries. 
(b) ($K_\textrm{s}$, $J$--$K_\textrm{s}$) DENIS CMD:
 298,928 entries. 
(c) Point sources present in 2MASS, DCMC and GSC2.2:
 1,161,701 entries. The $J$ and $K_\textrm{s}$ bands are from 2MASS.
(d) Point sources detected in 2MASS, DCMC but not GSC2.2:
 54,579 entries. The $J$ and $K_\textrm{s}$ bands are from 2MASS.
 (e) Point sources detected in 2MASS only: 151,120 entries.
(f) Point sources detected in both the IR catalogues and UCAC1: 
192,848 entries. The $J$ and $K_\textrm{s}$ bands are from 2MASS.
   } 
\end{figure*}

Figure~\ref{fig:comp}a shows the ($K_\textrm{s}$, $J$--$K_\textrm{s}$) diagram for all the 2MASS point sources.
The total number of sources, nearly two millions, was so large that we chose to plot them as isodensity curves, so as to emphasize
different loci of stars. 
The same technique has been adopted for most of the following diagrams.
Unfortunately, this process tends to hide regions with low density of stars. 
Sources in regions with density lower than the value
of the lowest contour level have been plotted as single dots.

The 2MASS colour--magnitude diagram (CMD) has been described in details by Nikolaev \& Weinberg (2000), and it will be taken as a reference for
the further discussion on the stellar populations obtained from the MC2.
Figure~\ref{fig:comp}b is a similar CMD, but for all the DCMC point sources.
Figure~\ref{fig:comp}c shows the CMD of the point sources that do have a counterpart in all three catalogues: DCMC, 2MASS and GSC2.2.
Figure~\ref{fig:comp}d shows the CMD of all the point sources detected in both
DCMC and 2MASS, but not GSC2.2. For Fig.~\ref{fig:comp}c and Fig.~\ref{fig:comp}d 
the $J$ and $K_\textrm{s}$ magnitudes are from 2MASS, including DCMC sources detected only in
$I$ and $J$.

{All the 2MASS sources that do not have any counterpart have been plotted on the CMD of Fig~\ref{fig:comp}e.
This feature is a mix of Asymptotic Giant Branch (AGB) and Red Giant Branch (RGB) stars.
The position of the AGB bump, located at the bottom of the AGB phase (see Gallart (1998) and references therein),
was found by Nikolaev \& Weinberg (2000) in the {\em deep} 2MASS observations at 
$K_\textrm{s}=15.8$ and ($J$--$K_\textrm{s}$)$=0.7$. 
The AGB bump stellar population has been well identified by Alcock et al. (2000) 
thanks to their 9 million LMC stars resulting from the MACHO project.
Note that Beaulieu \& Sackett (1998) call them the Supraclump.
The sensitivity limit is too low here to detect it, 
as for the red clump, which is located more than one magnitude below the AGB bump
($K_\textrm{s}\sim17$ and ($J$--$K_\textrm{s}$)$\sim 0.65$, Nikolaev \& Weinberg 2000).
}

Figure~\ref{fig:comp}f refers to sources detected in both 2MASS and UCAC1. 
{It shows mainly a concentration of stars around ($J$--$K_\textrm{s}$)$=0.5$ and 
$K_\textrm{s}=14$, which falls into
region D of Nikolaev \& Weinberg (2000). Note that Nikolaev \& Weinberg (2000) associate the blue half
part of region D with G--M dwarfs of the Galaxy. 
Ruphy et al. (1997) investigated the separation in ($J$--$K_\textrm{s}$) between dwarfs and giants, with the help
of early DENIS data, in the direction of the anticenter. They find that roughly for ($J$--$K_\textrm{s}$)$\leq0.6$ 
there could not be any giants. However, K and M dwarfs may be present for redder colours, together with the giants.
}
RGB stars at the tip of the RGB and AGB stars both O--rich and C--rich can be distinguished at ($J$--$K_\textrm{s}$)$\geq1.0$ 
(Cioni et al. 2000c).

\subsection{The ($J$--$H$, $H$--$K_\textrm{s}$) colour--colour diagram}
 
\begin{figure}
\begin{center}
		\psfig{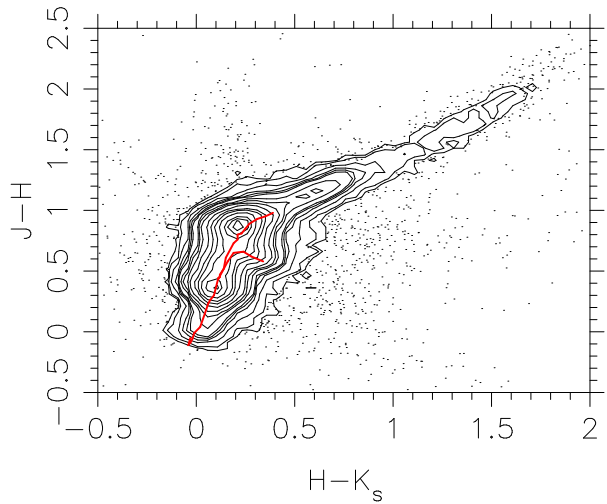} 
\end{center}
\caption[example] 
   { \label{fig:HK-JH}	  
Point sources detected in 2MASS, whatever the detection in the other catalogues is: 423,445 entries (photometric errors smaller than 0.06 magnitude).
The colour/colour dwarf and giant tracks are computed using Table 2 from Wainscoat et al. (1992).
 } 
\end{figure}

The ($J$--$H$, $H$--$K_\textrm{s}$) diagram may be used to discriminate between dwarf and giant stars,
at least to find the bifurcation between M dwarfs and M giants (Bessel \& Brett 1988). 
{Dwarfs are on the bluest peak, whereas giants are on the reddest one.}
This diagram contains only point sources with photometric errors on $J$,
$H$ and $K_\textrm{s}$ smaller than 0.06 magnitude.
{The ($J$--$H$, $H$--$K_\textrm{s}$) diagram is also suitable to isolate reddened stars. Nikolaev \& Weinberg (2000)
provide such a diagram and label the areas of unusual objects such as Wolf--Rayet stars, protostars, AGB C--rich stars and Be stars.
Frogel et al. (1990) surveyed several Magellanic Cloud clusters and overplotted on their resulting ($J$--$H$, $H$--$K_\textrm{s}$) diagram
the mean relations for globular cluster and field giants. The distribution of their cluster M giants is spread between these two lines, which
they relate to a metallicity effect. This could provide an explanation to the slight shift between the giant peak on Fig.~\ref{fig:HK-JH} and the 
giant track from Wainscoat et al. (1992) overplotted on it.}
{Finlator et al. (2000) cross--matched 2MASS with SDSS, thus selecting stars on their optical colours and
then tracing them in infrared diagrams. They found out that 
the dwarf peak in the ($J$--$H$, $H$--$K_\textrm{s}$) diagram is associated with stars earlier than G5, whereas the 
giant peak is associated with stars later than K5.
This is a typical example illustrating the usefulness of the combination of optical with infrared colours,
in order to separate stars according to their spectral type.
}

\subsection{IR/optical colour--colour diagrams}
\label{sect:colcol}

\begin{figure*}
\begin{center}
	\begin{tabular}{cc}
		\psfig{figure=JK-IJbis.ps,width=0.9\columnwidth,angle=-90}  & \psfig{figure=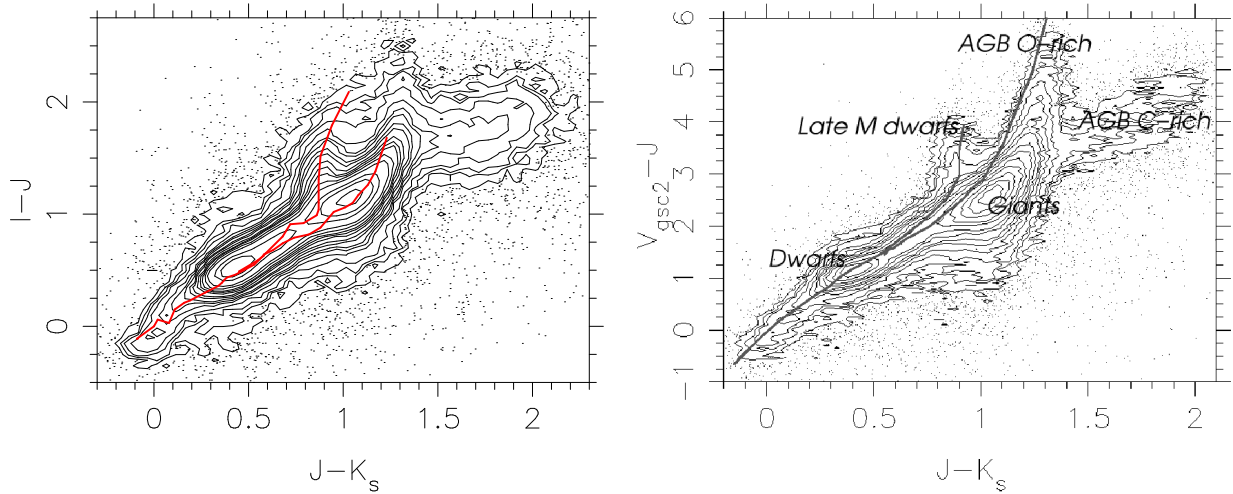,width=0.9\columnwidth,angle=-90} \\
	\end{tabular}
\end{center}
\caption[example] 
   { \label{fig:optical-IR}	  
Panel (a) contains sources detected by both DCMC and 2MASS: 372,354 entries. The $I$ band is from DENIS, whereas the $J$ and $K_\textrm{s}$ bands 
are from 2MASS. Dwarf and giant tracks superimposed are from Bessel \& Brett (1988).
Panel (b) contains sources detected by both GSC2.2 and 2MASS: 147,564 entries.
Dwarf and giant tracks are computed using Table 2 from Wainscoat et al. (1992).
The sources involved in those diagrams have photometric errors smaller than 0.06 magnitude.
   } 
\end{figure*}

Combining IR with optical wavelength, as shown on Fig.~\ref{fig:optical-IR}, 
enables us to discriminate between dwarf and giant stars.

{The two peaks show the combined effect of the fact that the contribution of the Galactic foreground stars are 
most likely due to the bluest dwarfs than to the reddest ones,
and that the limiting magnitude of the surveys excludes most LMC dwarfs. Otherwise, if all the populations of stars
were present, the two peaks would be merged.}
The separation between these two main clusters of stars is much better than in the ($J$--$H$, $H$--$K_\textrm{s}$) diagram.
Note that we plotted only sources with photometric errors on $I$, $J$ and $K_\textrm{s}$ smaller than 0.06 magnitude.
Two vertical sequences appear at ($J$--$K_\textrm{s}$)$=0.9$ (dwarfs) and ($J$--$K_\textrm{s}$)$=1.25$ (giants).
We identify the bluest vertical sequence with late M dwarfs, as suggested by the tracks superimposed on the 
($I$--$J$, $J$--$K_\textrm{s}$) and ($V$--$J$, $J$--$K_\textrm{s}$) diagrams.
{Note that the colour/colour giant track of both Wainscoat et al. (1992) and Bessel \& Brett (1988) 
do not exactly match the MC2 data. The shift is roughly 0.1 magnitude in ($J$--$K_\textrm{s}$), which could
be a photometric calibration problem. However it does not affect the track for the dwarfs which are mostly
galactic foreground stars.
As a consequence, since it affects only the track for the giants, it might be due
to metallicity or extinction effect.
}
{The search for late M, L and T dwarfs has been successful since the beginning of near--infrared sky surveys. 
But as pointed out by Leggett et al. (2002), infrared photometry alone does not allow to clearly
discriminate between the different spectral types. 
It is much easier to identify them on the basis of their optical/infrared colour index
(see also Kirkpatrick et al. 1999), because they are
so faint in the optical, and comparatively much brighter in the IR. These stars should disentangle 
themselves from the usual stars, and Reid et al. (2001) provide the location of some of these stars 
in the ($I$--$J$, $J$--$K_\textrm{s}$) CMD (and also ($J$--$H$, $H$--$K_\textrm{s}$)).
Smart et al. (2001) have stressed out the value of the GSC2 in the search for ultracool stars.
}

Some other well defined features (such as the M giant O--rich star and the C--star sequences) appear on each panel of 
Fig.~\ref{fig:optical-IR}, especially on the ($V$--$J$, $J$--$K_\textrm{s}$) diagram, 
where the spectral range between the optical and infrared magnitudes 
is much broader.

\subsection{The ($I$, $V$--$K_\textrm{s}$) colour--magnitude diagram}

\begin{figure}
\begin{center}
		\psfig{figure=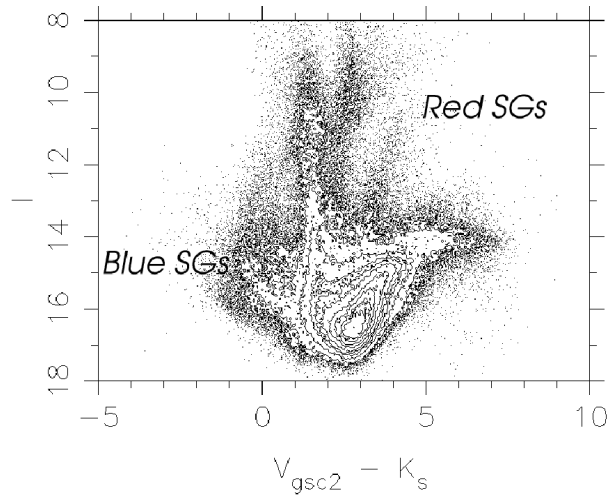,width=0.9\columnwidth,angle=-90} 
\end{center}
\caption[example] 
   { \label{fig:DG}	CMDs resulting from the cross--matching between
   the DCMC ($I$), 2MASS ($K_\textrm{s}$) and GSC2.2 ($V_{gsc2}$) catalogues. 
   This plot contains 393,179 entries.
   } 
\end{figure}

{We computed several CMDs using a combination 
of three different wavelengths, both IR and optical, out of the different catalogues.
The best features are obtained with the $V_{gsc2}$, $I$, and $K_\textrm{s}$ bands (Fig.~\ref{fig:DG}).
}

\begin{figure*}
\begin{center}
	\begin{tabular}{ccc}
		\psfig{figure=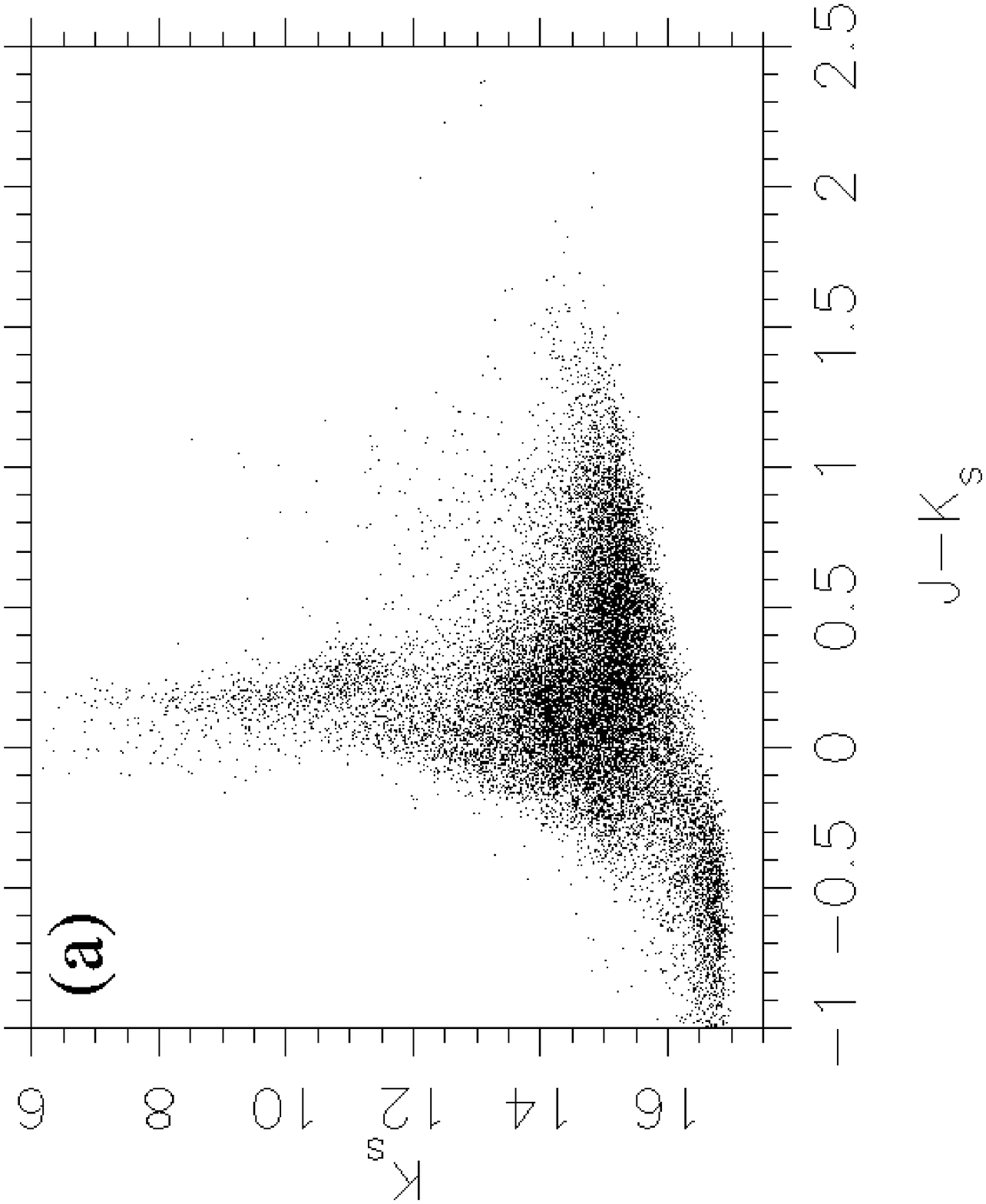,height=6cm,angle=-90} &
		\psfig{figure=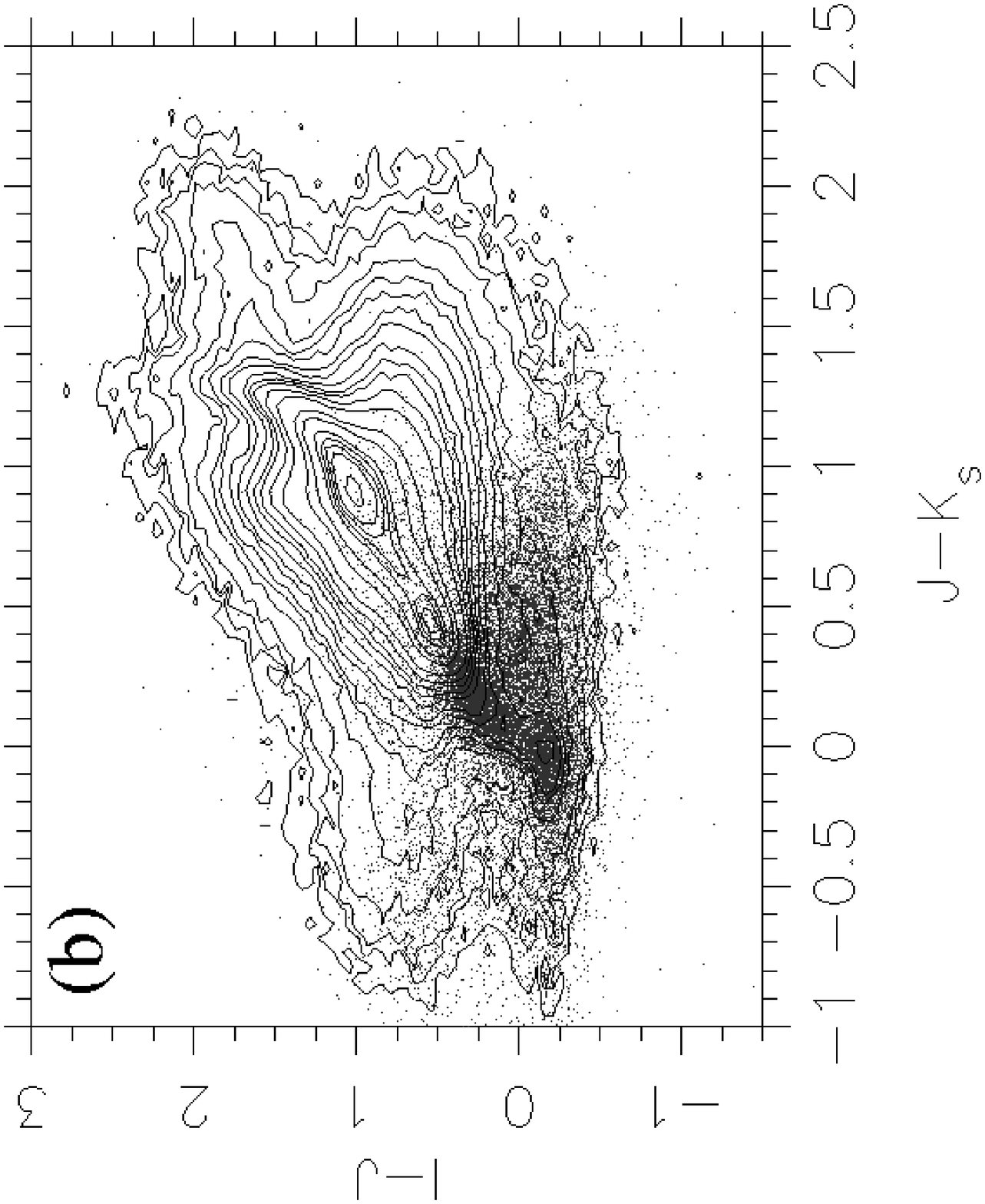,height=6cm,angle=-90} \\
		\psfig{figure=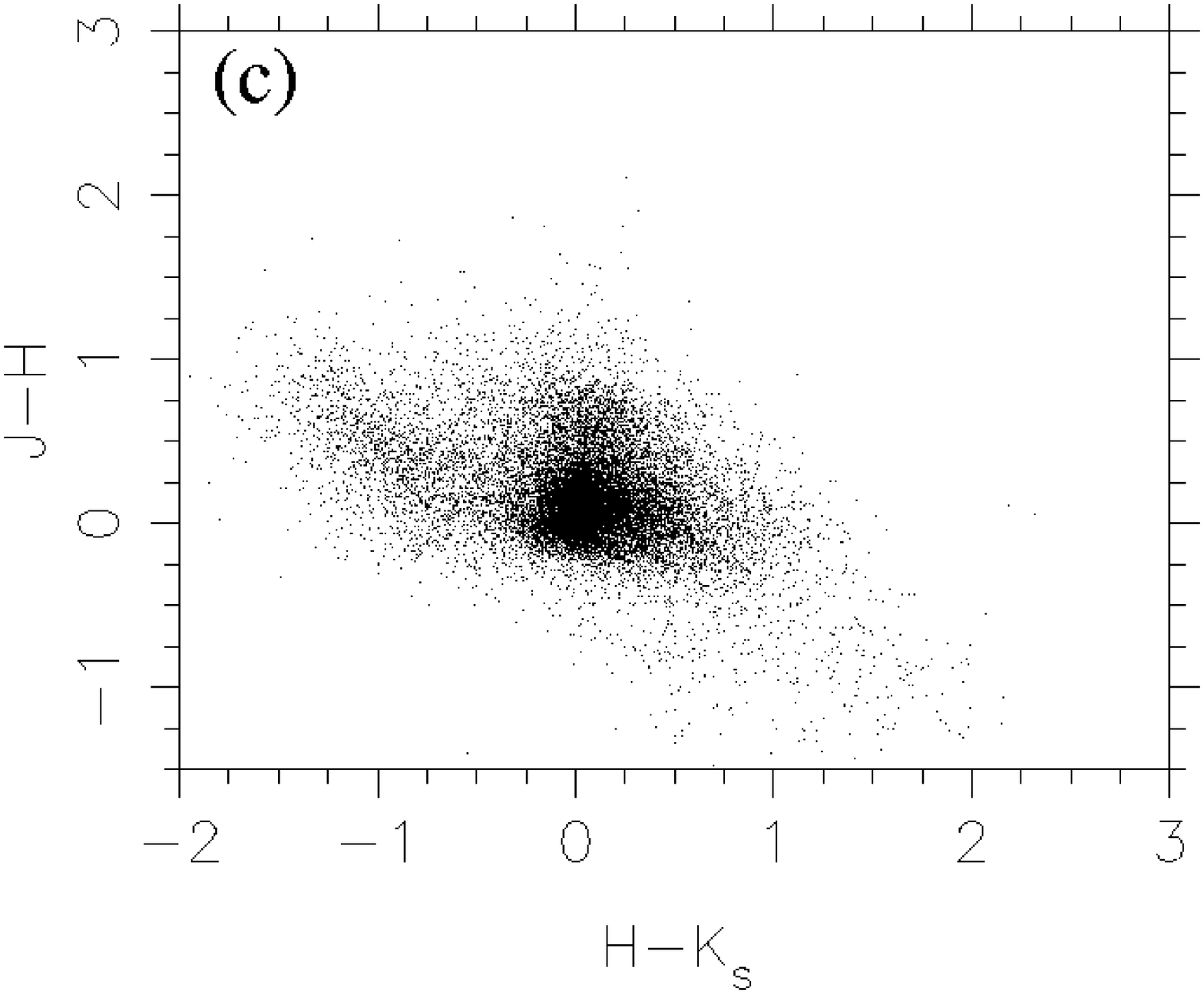,height=6cm,angle=-90} &
		\psfig{figure=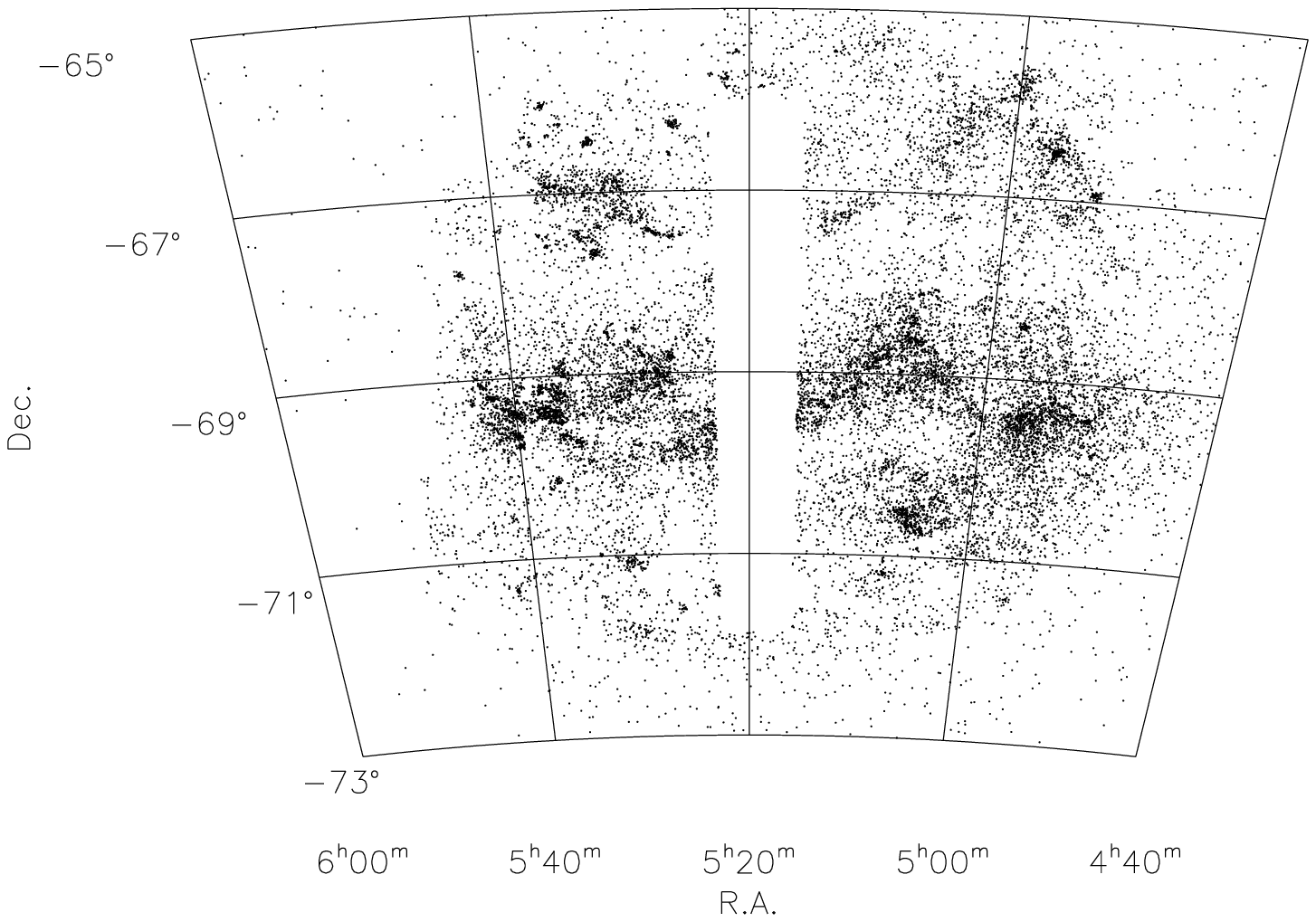,height=6cm} \\
	\end{tabular}
\end{center}
\caption[example] 
   { \label{fig:spiral}	 
   CMDs, colour--colour diagrams and spatial distribution of the blue stars selected with ($V$--$K_\textrm{s}$)$ \leq 1$ in the
   ($I$, $V$--$K_\textrm{s}$) diagram. 
   There are 19,646 sources on each panel. The $I$ band is from DENIS and the $J$, $H$ and $K_\textrm{s}$ are from 2MASS.
   The contour levels have been added so as to show the locus of different types of objects.
   } 
\end{figure*}

The red supergiants (SGs) are located in
the tight upward sequence at $I \sim 14$ and ($V$--$K_\textrm{s}$)$\sim 3$, while 
the blue SGs have ($V$--$K_\textrm{s}$)$ \leq 1$. 
This is consistent with the evolutionary tracks from Girardi et al. (2000).
We looked at the distribution of the stars with ($V$--$K_\textrm{s}$)$ \leq 1$
in various diagrams. The results are summarized on Fig~\ref{fig:spiral}.
They belong to the central parts of the LMC, and their spatial 
distribution is clumpy (Fig~\ref{fig:spiral}d), quite similar to what Martin et al. (1976) had found with their merging of several catalogues containing
SG stars. These sources are linked to the supergiant shells of the LMC (Meaburn 1980), which are probably produced by the effect
of stellar winds and/or supernovae.
These stars should help us constraining the recent star formation history of the LMC (Grebel \& Brandner 1998, Dolphin \& Hunter 1998).
Some of them fall into region A of Nikolaev \& Weinberg (2000) (Fig~\ref{fig:spiral}a): blue SGs, O dwarfs.
Since they are very blue stars, their ($V$--$K_\textrm{s}$) colour distinguish them from the bulk of stars on the
($I$, $V$--$K_\textrm{s}$) diagram (Fig~\ref{fig:DG}a,b).
They match the overdensity of stars at ($I$--$J$)$=-0.25$ and ($J$--$K_\textrm{s}$)$=0$ and extend towards redder colours (Fig~\ref{fig:spiral}b).
They are also recognizable on the ($J$--$H$, $H$--$K_\textrm{s}$) diagram at (0,0), at the bottom of the sequence of dwarfs (Fig~\ref{fig:spiral}c).
These young stars are much more easy to trace
in the IR/optical colour--colour diagrams and CMDs than in the ($K_\textrm{s}$, $J$--$K_\textrm{s}$) CMD.

\section{Conclusion}

The Master 
Catalogue of stars towards the Magellanic Clouds (MC2) is now available
on the web at CDS\footnote{\tt http://vizier.u-strasbg.fr/MC2/}.
It is a compilation of cross--identified surveys, from
optical to IR.
The MC2 roughly covers the following area: $4^{h}$ to $7^{h}$ in Right Ascension,
and $-61^{\circ}$ to $-78^{\circ}$ in Declination, with slight variations
according to the catalogue considered.

We are currently working on the cross--identification of analogous catalogues in the direction of the Small
Magellanic Cloud and we plan to add 
catalogues and tables at other wavelengths: ROSAT, IRAS,and many more
specific catalogues, as well the variability informations coming either from MACHO, EROS, OGLE to the second version of the catalogue.

A typical query of the MC2 returns several lines of data.
{An example is given in Table~3.}
Each line of data contains the name of the source for all the original catalogues, followed
by the magnitudes and the proper motion when the UCAC1 is present. 
For each catalogue, the distance 
of the cross--identification is given, except for 2MASS which is taken as reference.
The distance associated to a DCMC source is the distance to the 2MASS counterpart. 
The distance associated to a GSC2.2 source is the distance to the 2MASS counterpart, or the DCMC counterpart
when there is no 2MASS counterpart.
The distance associated to a UCAC1 source is the distance to the MC2
counterpart (2MASS, DCMC or GSC2.2, depending on the detection bands).
At the beginning of each line, R.A. and Dec. are given. The choice of the coordinates has been as follows: 
when possible, we kept the 2MASS coordinates as the reference, otherwise we took the GSC2.2, then the UCAC1 and 
finally the DCMC ones. 
\begin{table*} [h]   
\small
\begin{center}
	\caption{Subsample of the MC2. Each line of data corresponds to one point source.
        }

\begin{tabular}{cccccccccccccccccccccccccccccc}
   \hline
R.A.     &     Dec. &        2MASS &   J  &  dJ &  H   & dH  &   Ks & dKs \\
   \hline
93.769831&-75.633698&0615047-753801&12.380&0.022&12.177&0.024&12.076&0.029\\
93.760565&-75.632538&0615025-753757&16.072&0.089&15.975&0.178&15.273&null \\
93.769231&-75.629761&0615046-753747&16.413&0.123&16.141&0.211&15.611&0.261\\
93.756995&-75.621536&0615016-753717&16.192&0.098&15.352&0.106&15.443&0.215\\
   \hline
       DCMC	  &	I&  dJ &   J  &   dJ&	Ks &  dKs &  dist  \\
   \hline
061504.60-753800.7&12.999&0.006&15.788&0.188&99.000&99.000&0.853030\\
061502.40-753757.5&16.501&0.062&15.866&0.230&99.000&99.000&0.621811\\
..		  &...   &...  &...   &...  &...   &...   &...     \\
..		  &...   &...  &...   &...  &...   &...   &...     \\
   \hline
     GSC2.2 &F    &dF       &J    &dJ       &V   &dV   &dist    \\
   \hline
S1102121266 &13.24&$\pm$0.23&13.95&$\pm$0.17&--- &$\pm$&0.192289\\
S11021214338&17.40&$\pm$0.23&17.90&$\pm$0.18&--- &$\pm$&0.203876\\
S11021214339&17.44&$\pm$0.23&18.29&$\pm$0.18&--- &$\pm$&0.554134\\
S11021214495&18.30&$\pm$0.24&---  &$\pm$    &--- &$\pm$&0.251949\\
   \hline
	UCAC1	&mag  &PMra &PMdec& dist   \\
   \hline
00697288&13.22&+27.3&-9.7 &0.045439\\
...	&...  &...  &...  &...     \\
...	&...  &...  &...  &...     \\
...	&...  &...  &...  &...     \\
\end{tabular}							

\end{center}
\end{table*}

We decided to keep the distances of the cross--identifications in the MC2 to give the user
the opportunity to judge the reliability of each cross--identification. We have also
shown that for some DCMC point sources, this distance was not reliable.
But since links for each source allow to access the complete data from
the original catalogues through the VizieR search engine (Ochsenbein et al.\ 2000),
it is always possible to retrieve the strip number of the DCMC source and then go to the MC2
web site to find the shifts associated to this strip.
Those links are also very valuable in order to retrieve observational data such as image or scan
number, flags or whatever parameter the user would like to know from the original catalogues.

This reference catalogue is made available as a support for a number of studies 
concerning, e.g. the stellar populations in the Magellanic Clouds, the structure of the 
Clouds, or certain classes of objects (Cepheids, AGB stars, etc.).  
{ Recent articles, 
such as those by Zaritsky et al. (2002), Van der Marel (2001), Nikolaev \& Weinberg (2001) and 
Cioni et al.\ (2000b) have 
demonstrated the power of optical and near--infrared surveys to improve our understanding on these
neighbouring galaxies.}

%

\begin{acknowledgements}
We thank an anonymous referee for very constructive suggestions which helped
  to improve the present paper.
We would like to thank Fran\c cois Ochsenbein for his help with the catalogues.
This research has made use of the SIMBAD astronomical database,
  the VizieR catalogue service, and the ALADIN interactive sky atlas,
  all operated at CDS, Strasbourg, France.
This work has been partly supported
    by the {\sc astrovirtel} Project which is run by the ESO/ST-ECF
    Archive and funded by the European Commission under contract
    HPRI-CT-1999-00081.
This publication makes use of data products from 
   the Two Micron All Sky Survey, which is
   a joint project of the University of Massachusetts and the Infrared Processing and Analysis
   Center/California Institute of Technology, funded by the National Aeronautics and Space
   Administration and the National Science Foundation, and from 
   DENIS, 
   which is the result of a joint effort involving human and financial contributions of 
   several Institutes mostly located in Europe. It has been supported financially mainly 
   by the French Institut National des Sciences de l'Univers, CNRS, and French 
   Education Ministry, the European Southern Observatory, the State of 
   en-W\"urttemberg, 
   and the European Commission under a network of the Human Capital and Mobility program.
The Guide Star Catalogue-II is a joint project of the Space Telescope Science Institute and the Osservatorio Astronomico di
   Torino. Space Telescope Science Institute is operated by the Association of Universities for Research in Astronomy, for the National
   Aeronautics and Space Administration under contract NAS5-26555. The participation of the Osservatorio Astronomico di Torino is
   supported by the Italian Council for Research in Astronomy. Additional support is provided by European Southern Observatory, Space
   Telescope European Coordinating Facility, the International GEMINI project and the European Space Agency Astrophysics Division.
\end{acknowledgements}


\end{document}